\documentclass[aps,prd,twocolumn,superscriptaddress,nofootinbib,eqsecnum]{revtex4}

\pdfoutput=1

\usepackage{amsfonts}
\usepackage{amsmath}
\usepackage{amssymb}
\usepackage{graphicx,color}
\usepackage{float}
\usepackage{hyperref}
\usepackage{subfigure}
\usepackage{dcolumn}% Align table columns on decimal point

\begin{document}

\title{Regularization issues for a cold and dense quark matter model
  in $\beta$ equilibrium}

\author{Dyana C. Duarte}
\email{dyana@ita.br}
\affiliation{Departamento de F\'{\i}sica, Instituto Tecnol\'ogico de
  Aeron\'autica, 
12228-900 S\~ao Jos\'e dos Campos, SP, Brazil}

\author{R. L. S. Farias}
\email{ricardo.farias@ufsm.br}
\affiliation{Departamento de F\'{\i}sica, Universidade Federal de Santa
  Maria, 97105-900 Santa Maria, RS, Brazil}

\author{Rudnei O. Ramos}
\email{rudnei@uerj.br}
\affiliation{Departamento de F\'{\i}sica Te\'orica, Universidade do
  Estado do Rio de
Janeiro, 20550-013 Rio de Janeiro, RJ, Brazil}

\begin{abstract}

A regularization scheme that explicitly separates vacuum contributions
from medium effects is applied to a  Nambu--Jona-Lasinio model with
diquark interactions and in $\beta$ equilibrium.  We perform a
comparison of this proposed scheme with the more traditional one,
where no separation of vacuum and medium effects is done. Our results
point to both qualitative and quantitative important differences
between these two methods, in particular regarding the phase structure of
the model in the cold and dense nuclear matter case. 

\end{abstract}

\maketitle

%%%%%%%%%%%%%%%%%%%%%%%%%%%%%%%%%%%%%%%%%%%%%%%%%%%%%%%%%%%%%%%%%%%%
\section{Introduction}

Low-energy nonrenormalizable effective models are widely used as a
tool to understand many  problems in physics when its microscopic,
renormalizable counterpart is too  complex to be used. This is
particularly true in the case of quantum chromodynamics (QCD), 
where a full use of it becomes not applicable in the
context of perturbation theory, like at low energies, given its strong
coupling nature. Likewise, the study of the dense nuclear matter
through nonperturbative methods based on lattice Monte Carlo QCD
simulations (for a recent review, see, e.g., Ref.~\cite{Ding:2015ona}
and references therein) is plagued by the so-called  ``sign
problem.'' In this context, the use of effective models, for instance
the Nambu--Jona-Lasinio (NJL) type of models~\cite{KLEVAN,BUBA}, is
highly valuable  to access some otherwise unaccessible region of
parameters.  

Even in the context of effective models, we still find ultraviolet
(UV) divergent momentum integrals that need to be solved.  The usual
prescription adopted in the literature is to regularize all the
divergent momentum integrals through a sharp cutoff $\Lambda$. The
momentum cutoff $\Lambda$ in this case is then treated as a parameter
of the model, which is fitted by the physical quantities (e.g., by
using the pion decay constant, the quark condensate, and pion mass).
$\Lambda$ also in a sense sets an energy scale below which in general
the effective  model should be trusted. 

In Refs.~\cite{Alford1998,Rapp1998} the authors showed
that the gaps in fermion spectrum  are expected to be of the order of
100 MeV and that the possible presence of color superconducting
phases could be extended  to the region of nonzero temperature on
the QCD phase diagram. This is related to the fact that,  just like
in the case of the usual superconductivity, gaps are related to
larger values of critical  temperature, which results in a very rich
phase structure. In this region the matter consists of three quarks
and, depending on the value of the strange ($s$) quark mass, one may
observe different types of superconducting phases that can be
formed by one,  two, or three flavors of quarks. If three quarks do
participate in the pairing, the color-flavor  locked (CFL) phase is
observed~\cite{Alford:1998mk}, while if only the up ($u$) and down
($d$) quarks take  place in the pairing, the phase is two color superconductor 
(2SC) + $s$. It is also possible that only the $s$ quark forms pairs,
characterizing a spin-1 condensate that may have influence on some
properties of compact stars,  as shown, e.g.,  by
Ref.~\cite{Schmitt:2003xq}. If the value of the baryon density is
not large enough  such that the quark $s$ does not need to be
included, we have a pairing of up and down quarks and the
corresponding phase is called 2SC. This is the case we are
interested in the present work.
In Refs.~\cite{Rajagopal:2000wf,Alford:2001dt,Rischke:2003mt} there
are good  reviews on this topic, and in~\cite{Son1999,Shovkovy1999}
the authors studied the color superconductivity  mechanism at
asymptotic densities using first principles calculations. One of
the most relevant applications of these studies is to understand the
structure  of compact stars, where color superconducting phases are
expected, since these stars are expected to present densities  on
their nuclei of the order of 10 $\rho_0$, where $\rho_0 \sim 0.15$
fm$^{-3}$ is the saturation density.

In the present work we will make use of the NJL model to study the
diquark condensation for  a cold and dense quark matter with color and
electric charge neutrality.  The importance of this model stems from
the fact that at sufficiently cold and dense regimes,  quark matter
behaves as a color superconductor~\cite{Barrois1977,Bailin1984}, where
quarks  form Cooper pairs with equal and opposite momenta, and
studies of QCD under these conditions  have many different
applications.  When studied in the context of the NJL model, which is
an intrinsically  nonrenormalizable model, we should in principle
handle with care the regularization procedure.  Even though the
regularization scheme is treated as part of the model, it can
potentially mix vacuum  quantities with medium ones, which can involve
explicitly, e.g., chemical potentials, temperature and external fields,
or implicilty, through, e.g., a dependence on the various condensates
that the system can allow.  In principle, one could claim that we are
not restricted to follow any prescribed  regularization procedure, but
we still find that is fair to assume that the model, 
since it is supposed to be an approximation of a
renormalizable one (QCD), hence we believe that preserving those
properties observed in the latter case is desirable.  One of these
properties is that renormalization should in principle only depend on
vacuum quantities and not on medium effects. Thus, regularization
procedures are required to be done only on vacuum dependent terms,
while medium dependent terms should be independent of the
regularization chosen to perform UV divergent terms. 
That this separation of medium effects from regularized vacuum
dependent terms only might have not only qualitative but also
important quantitative effects was noticed already in
Ref.~\cite{Farias2006} in the context of color superconductivity in a
NJL model.  A similar situation was also found in the case of studies of
magnetized quark matter, where unphysical spurious effects are
eliminated by properly separating the magnetic field contributions
from  the divergent
integrals~\cite{Menezes:2008qt,Allen2015,Duarte:2015ppa,Coppola2017}. This
same procedure has been advocated recently in
Ref.~\cite{Farias:2016let} in the context of the NJL model with a
chiral imbalance. The process of disentangling the  vacuum dependent
terms from medium ones and properly regularizing only the UV divergent
momentum integrals for the former was named the ``medium separation scheme'' 
(MSS) in opposition to the usual procedure where the cutoff
is applied even to the medium dependent terms, the {\it traditional
regularization  scheme} (TRS). 

One of the most important motivations for our study comes from the fact
that there is a great deal of evidence pointing to the increasing of the
diquark condensate with the chemical potential. While realistic
$N_{c}=3$ QCD lattice simulations cannot be implemented due to the
well-discussed sign problem~\cite{Karsch2002} this is not the case
for $N_{c}=2$. Using two colors, the lattice simulations are
available, and the results clearly predicts an increase of the diquark condensate 
with the chemical potential~\cite{Kogut2001,Braguta2016}. This
is also indicated by studies using chiral perturbation theory
(ChPT)~\cite{Kogut:2000ek}.   As we are going to see, in the
traditional treatment of divergences in the TRS case, the diquark
condensate eventually vanishes, which seems to be at odds with what we
would expect in general. By applying the MSS procedure instead, we
predict an always  increasing diquark condensate. As a side result, we
also show an explicit change of behavior in the phases for the diquark
condensate as the chemical potential is increased, which is not able
to be obtained in the context of the TRS regularization procedure.

This work is organized as follows. In Sec.~\ref{sec2}, we introduce
the NJL model with diquark interactions and give the explicit
expression for the thermodynamical potential. In Sec.~\ref{sec3} we
give the MSS regularization  procedure and define the appropriate
equations for having $\beta$ equilibrium  and charge neutrality.  In
Sec.~\ref{results}, we show our numerical results comparing TRS and
MSS schemes. In Sec.~\ref{conclusions}, we present our conclusions.
Three appendixes are also included where we show the more technical
details in the MSS calculation and how to write the
expressions for the individual  densities when $\beta$ equilibrium
is included.

%%%%%%%%%%%%%%%%%%%%%%%%%%%%%%%%%%%%%%%%%%%%%%%%%%%%%%%%
\section{The model and its thermodynamic potential}
\label{sec2}

In this work we study the NJL model with interactions involving the
scalar, pseudoscalar and diquark channels for the quark field.  The
explicit form of the Lagrangian density is then given by
\begin{eqnarray}
{\cal L} & = &
\bar{\psi}\left(i\gamma^{\mu}\partial_{\mu}-m\right)\psi+G_{s}
\left[\left(\bar{\psi}\psi\right)^{2}
  +\left(\bar{\psi}i\gamma_{5}\vec{\tau}\psi\right)^{2}\right]\nonumber
\\ &  & +
G_{d}\left[\left(i\bar{\psi}^{C}\varepsilon\epsilon^{b}\gamma_{5}\psi\right)
  \left(i\bar{\psi}\varepsilon\epsilon^{b}\gamma_{5}\psi^{C}\right)\right],
\label{lag}
\end{eqnarray}
where $m$ is the current quark mass, $\psi^{C}=C\bar{\psi}^{T}$ is the
charge-conjugate spinor, and $C=i\gamma^{2}\gamma^{0}$ is the charge
conjugation matrix. The quark field $\psi\equiv\psi_{i\alpha}$ is a
four-component Dirac spinor that carries both flavor
$\left(i=1,2\right)$ and color $\left(\alpha=1,2,3\right)$
indices. The Pauli matrices are denoted by
$\vec{\tau}=\left(\tau^{1},\tau^{2},\tau^{3}\right)$, while
$\left(\varepsilon\right)^{ik}\equiv\varepsilon^{ik}$ and
$\left(\epsilon^{b}\right)^{\alpha\beta}\equiv\epsilon^{\alpha\beta
  b}$ are the antisymmetric tensors in the flavor and color spaces,
respectively.

In $\beta$ equilibrium, the diagonal matrix of quark chemical
potentials is given in terms of quark, electrical charge, and color charge
chemical potentials,
\begin{equation}
\mu_{ij,\alpha,\beta}=\left(\mu\delta_{ij}-\mu_{e}Q_{ij}\right)\delta_{\alpha\beta}
+\frac{2}{\sqrt{3}}\mu_{8}\delta_{ij}\left(T_{8}\right)_{\alpha\beta},
\end{equation}
where $Q$ and $T_{8}$ are generators of the electromagnetism
U(1)$_{\text{em}}$ and the U(1)$_{8}$ subgroup of the color gauge
groups, respectively.  The explicit expressions for the quark chemical
potentials read 
\begin{eqnarray}
\mu_{ur} & = &
\mu_{ug}=\mu-\frac{2}{3}\mu_{e}+\frac{1}{3}\mu_{8},\label{muur}\\ \mu_{dr}
& = &
\mu_{dg}=\mu+\frac{1}{3}\mu_{e}+\frac{1}{3}\mu_{8},\label{mudr}\\ \mu_{ub}
& = &
\mu-\frac{2}{3}\mu_{e}-\frac{2}{3}\mu_{8},\label{muub}\\ \mu_{db} & =
& \mu+\frac{1}{3}\mu_{e}-\frac{2}{3}\mu_{8}.\label{mudb}
\end{eqnarray}

In the mean field approximation, the finite temperature ($T$)
effective potential for quark matter in $\beta$ equilibrium with
electrons is well known~\cite{Huang:2003xd,Shovkovy:2003uu} and it is
explicitly given by
\begin{eqnarray}
\Omega & = &
\Omega_{0}-\left(\frac{\mu_{e}^{4}}{12\pi^{2}}+\frac{T^{2}\mu_{e}^{2}}{6}
+\frac{7\pi^{2}}{180}T^{4}\right)+\frac{\left(M-m\right)^{2}}{4G_{s}}
\nonumber \\ &+&\frac{\Delta^{2}}{4G_{d}} -  2
T\sum_{a}n_{a}\int\frac{d^{3}p}{\left(2\pi\right)^{3}} \ln\left(1+e^{-
  E_{a}/T}\right) \nonumber
\\ &-&\sum_{a}n_{a}\int\frac{d^{3}p}{\left(2\pi\right)^{3}}E_{a},
 \label{OmegaT}
\end{eqnarray}
where $\mu_{e}$ is the electron chemical potential and $\Omega_{0}$ is
the constant vacuum energy term added so as to make the pressure  of the
vacuum zero.  In Eq.~(\ref{OmegaT}), for simplicity, we have assumed a
vanishing value for  the electron mass, which is sufficient for the
purposes of the current study.  The sum in the second line of
Eq.~(\ref{OmegaT}) runs over all quasiparticles, 
whose explicit dispersion relations read
\begin{eqnarray}
E_{ub}^{\pm} & = & E\pm\mu_{ub},\label{Eub}\\ E_{db}^{\pm} & = &
E\pm\mu_{db},\label{Edb}\\ E_{\Delta^{\pm}}^{\pm} & = &
E_{\Delta}^{\pm}+\delta\mu,\label{Erg}
\end{eqnarray}
where we have introduced the following notation for convenience:
\begin{eqnarray}
  E & \equiv & \sqrt{p^{2}+M^{2}},\label{E}
  \\
E_{\Delta}^{\pm} & \equiv
&
  \sqrt{\left(E\pm\bar{\mu}\right)^{2}+\Delta^{2}},\label{Ed}
  \\
  \bar{\mu}
& \equiv & \frac{\mu_{ur}+\mu_{dg}}{2}=\frac{\mu_{ug}+\mu_{dr}}{2}
  \nonumber \\
  &=&  \mu-\frac{\mu_{e}}{6}
+\frac{\mu_{8}}{3},\label{mubar}\\ \delta\mu & \equiv &
\frac{\mu_{dg}-\mu_{ur}}{2}=\frac{\mu_{dr}+\mu_{ug}}{2}
\nonumber
\\ &=&\frac{\mu_{e}}{2};
\label{dmu}
\end{eqnarray}
in the above equations, $\Delta$ is the diquark condensate 
and $M$ is the constituent quark mass. The
multiplicity $n_{a}$ in Eq.~(\ref{OmegaT}) is related to the
degeneracy factors of each quasiparticle dispersions, such as
$n_{ub}=n_{db}=1$ and $n_{\Delta}=2$ (corresponding to the
dispersions $E_{\Delta^{\pm}}^{\pm}$, related to the $r$ and $g$
colors, due to the definitions of $\bar{\mu}$ and $\delta\mu$).

{}For the demonstration purposes in this work, we can assume, without
loss of generality,  the chirally symmetric phase of quark matter and,
thus, we will work in the chiral limit.  In 
this limit, the quasiparticle dispersions~(\ref{E})
and (\ref{Ed}) then become
\begin{eqnarray}
E & = & p,\label{Ep}\\ E_{\Delta}^{\pm} & = &
\sqrt{\left(p\pm\bar{\mu}\right)^{2}+\Delta^{2}}.\label{Edelta}
\end{eqnarray}
The vacuum term $\Omega_{0}$ in Eq.~(\ref{OmegaT}) is obtained by
considering $\Delta=\mu=\mu_{e}=\mu_{8}=0$ and taking the effective
quark mass is its vacuum value,  $M_{0}$. Thus, we obtain that
\begin{eqnarray}
\Omega_{0} & = &
-\frac{M_{0}^{2}}{4G_{s}}+12\int\frac{d^{3}p}{\left(2\pi\right)^{3}}
\sqrt{p^{2}+M_{0}^{2}}.
\label{Omega0}
\end{eqnarray}
{}Finally, by considering the $T\to 0$ limit in Eq.~(\ref{OmegaT}), we
obtain that~\cite{Huang:2003xd}
\begin{eqnarray}
\Omega_{T=0}\left(\Delta,\bar{\mu},\delta\mu\right)  & = &
\Omega_{0}-\frac{\mu_{e}^{4}}{12\pi^{2}}+\frac{\Delta^{2}}{4G_{d}} -
\frac{\mu_{ub}^{4}}{12\pi^{2}}-\frac{\mu_{db}^{4}}{12\pi^{2}}
\nonumber \\ & - &
2\theta\left(\delta\mu-\Delta\right)\int_{\mu^{-}}^{\mu^{+}}
\frac{dp}{\pi^{2}}p^{2}\left(\delta\mu-E_{\Delta}^{-}\right) \nonumber
\\ &-&
4\int\frac{d^{3}p}{\left(2\pi\right)^{3}}\left(p+E_{\Delta}^{+}+
E_{\Delta}^{-}\right),
 \label{OmegaT0m0}
\end{eqnarray}
where $\mu^{\pm}=\bar{\mu}\pm\sqrt{\delta\mu^{2}-\Delta^{2}}$ and
$\Omega_{0}$ is given by Eq.~(\ref{Omega0}).

%%%%%%%%%%%%%%%%%%%%%%%%%%%%%%%%%%%%%%%%%%%%%%%%%%%%%%%%%%%%%%%%%%%%
\section{Regularization issues and medium effects}
\label{sec3}

The momentum integral in the last term in Eq.~(\ref{OmegaT}) and,
equivalently, the last one  in Eq.~(\ref{OmegaT0m0}) when taking the
chiral limit are UV divergent.  These terms mix vacuum quantities
with medium ones. In the case of the last term in Eq.~(\ref{OmegaT}),
or in the case of Eq.~(\ref{OmegaT0m0}), we have contributions that depends
explicitly or implicilty in chemical potentials $\mu$, $\mu_e$, and $\mu_8$,
and in the diquark condensate $\Delta$. Integrands with these dependences 
might not be regularized naively, just introducing a cutoff parameter.
As argued in the Introduction, these terms should be handled with care
and two regularization  procedures can be applied, the TRS and the MSS
one.  Let us start by discussing the implementation of the MSS
procedure as a way to properly disentangle these dependencies of vacuum
dependent terms from the medium effects for the present problem.  We
initially discuss the more general case, the physical case, with a
nonvanishing current quark mass $m$. The generalization for the chiral
limit for the MSS regularized integrals is immediate.

The gap equation for $\Delta$ is obtained from Eq.~(\ref{OmegaT}) by
deriving it with respect to $\Delta$. Let us concentrate  on the UV
divergent term that results from the last term in Eq.~(\ref{OmegaT}).
Its contribution for the gap equation for $\Delta$ is of the form
\begin{eqnarray}
I_{\Delta} & = &
\int\frac{d^{3}p}{\left(2\pi\right)^{3}}\left(\frac{1}{E_{\Delta}^{+}}
+\frac{1}{E_{\Delta}^{-}}\right) \nonumber \\ & = &
\sum_{s=\pm1}\int\frac{d^{3}p}{\left(2\pi\right)^{3}}\frac{1}{E_{\Delta}^{s}}\,,
 \label{tempIvD}
\end{eqnarray}
with $E_{\Delta}^{s}=\sqrt{\left(E+s\bar{\mu}\right)^{2}+\Delta^{2}}$.
This term can be rewritten as 
\begin{equation}
\int\frac{d^{3}p}{\left(2\pi\right)^{3}}\frac{1}{E_{\Delta}^{s}}=\frac{1}{\pi}
\int_{-\infty}^{+\infty}dp_{4}\int\frac{d^{3}p}{\left(2\pi\right)^{3}}
\frac{1}{p_{4}^{2}+(E_{\Delta}^{s})^{2}}\,,
\end{equation}
such that 
\begin{eqnarray}
\lefteqn{\frac{1}{2}
  \sum_{s=\pm1}\int\frac{d^{3}p}{\left(2\pi\right)^{3}}
  \frac{1}{E_{\Delta}^{s}} } \nonumber \\ &&
=\sum_{s=\pm1}\int_{-\infty}^{+\infty}
\frac{dp_{4}}{2\pi}\int\frac{d^{3}p}{\left(2\pi\right)^{3}}
\frac{1}{p_{4}^{2}+(E_{\Delta}^{s})^{2}}.
\label{identD4k}
\end{eqnarray}
Making use of the identity~\cite{Battistel1999}, 
\begin{eqnarray}
\lefteqn{ \frac{1}{p_{4}^{2}+\left(E+s\bar{\mu}\right)^{2}+\Delta^{2}}
} \nonumber \\ & & =  \frac{1}{p_{4}^{2}+p^{2}+M_{0}^{2}} \nonumber
\\ & & -
\frac{\mu^{2}+2sE\mu+\Delta^{2}+M^{2}-M_{0}^{2}}{\left(p_{4}^{2}+E^{2}+
  M_{0}^{2}\right)
  \left[p_{4}^{2} +\left(E+s\bar{\mu}\right)^{2}+\Delta^{2}\right]},
 \label{ident}
\end{eqnarray}
we obtain, after making two iterations of this same identity, the
result
\begin{eqnarray}
\lefteqn{ \frac{1}{p_{4}^{2}+\left(E+s\bar{\mu}\right)^{2}+\Delta^{2}}
} \nonumber \\ & & =  \frac{1}{p_{4}^{2}+p^{2}+M_{0}^{2}}
+\frac{A-2sE\bar{\mu}}{\left(p_{4}^{2}+p^{2}+M_{0}^{2}\right)^{2}}
\nonumber \\ &  & +
\frac{\left(A-2sE\bar{\mu}\right)^{2}}{\left(p_{4}^{2}+p^{2}+M_{0}^{2}\right)^{3}}
\nonumber \\ & & +
\frac{\left(A-2sE\bar{\mu}\right)^{3}}{\left(p_{4}^{2}+
  p^{2}+M_{0}^{2}\right)^{3}\left[p_{4}^{2}
    +\left(E+s\bar{\mu}\right)^{2}+\Delta^{2}\right]},
 \label{r3it}
\end{eqnarray}
with $A=M_{0}^{2}-M^{2}-\bar{\mu}^{2}-\Delta^{2}$. Thus, after
performing some simple algebraic manipulations, the sum in $s$ and
also the $p_{4}$ integrations, as indicated in Eq.~(\ref{identD4k}),
can be made and we obtain  the result
\begin{eqnarray}
\lefteqn{ \sum_{s=\pm1} \int_{-\infty}^{+\infty}\frac{dp_{4}}{2\pi}
  \int\frac{d^{3}p}{\left(2\pi\right)^{3}}\frac{1}{p_{4}^{2}+(E_{\Delta}^{\pm})^{2}}
} \nonumber \\ & & =
I_{\text{quad}}-\frac{\left(\Delta^{2}-M_{0}^{2}-2\bar{\mu}^{2}+M^{2}\right)}{2}I_{\text{log}}
-  I_{\text{fin,2}}\nonumber \\ &&  \nonumber \\ & & +
\left[\frac{3\left(A^{2}+4M^{2}\bar{\mu}^{2}\right)}{8}
  -\frac{3\bar{\mu}^{2}M_{0}^{2}}{2}\right]I_{\text{fin,1}} ,
 \label{temp4}
\end{eqnarray}
where we have defined the quantities
\begin{eqnarray}
I_{\text{quad}} & = &
\int\frac{d^{3}p}{\left(2\pi\right)^{3}}\frac{1}{\sqrt{p^{2}+M_{0}^{2}}},
\label{quad}
\\ I_{\text{log}} & = & \int\frac{d^{3}p}{\left(2\pi\right)^{3}}
\frac{1}{\left(p^{2}+M_{0}^{2}\right)^{\frac{3}{2}}},
\label{log}
\\ I_{\text{fin,1}} & = & \int\frac{d^{3}p}{\left(2\pi\right)^{3}}
\frac{1}{\left(p^{2}+M_{0}^{2}\right)^{\frac{5}{2}}},
\label{Ifin1}
\\ I_{\text{fin,2}} & = &
\frac{15}{32}\sum_{s=\pm1}\int\frac{d^{3}p}{\left(2\pi\right)^{3}}
\int_{0}^{1} dx (1-x)^{2} \nonumber \\ &\times&
\frac{(A-2sE\bar{\mu})^{3}}
     {\left[(2sE\bar{\mu}-A)x+p^{2}+M_{0}^{2}\right]^{\frac{7}{2}}}.
 \label{Ifin2}
\end{eqnarray}

Comparing Eq.~(\ref{tempIvD}) and the left-hand side of
Eq.~(\ref{identD4k}) we can see that
\begin{equation}
\frac{1}{2}I_{\Delta}=\sum_{s=\pm1}\int_{-\infty}^{+\infty}\frac{dp_{4}}{2\pi}
\int\frac{d^{3}p}{\left(2\pi\right)^{3}}\frac{1}{p_{4}^{2}+
  (E_{\Delta}^{\pm})^{2}},
\end{equation}
and, therefore,
\begin{eqnarray}
I_{\Delta} & = & 2I_{\text{quad}}
-\left(\Delta^{2}-M_{0}^{2}-2\bar{\mu}^{2}+M^{2}\right)
I{}_{\text{log}}+2I_{\text{fin,2}}\nonumber \\ & + &
\left[\frac{3(A^{2}+4M^{2}\bar{\mu}^{2})}{4}-3M_{0}^{2}\bar{\mu}^{2}\right]
I_{\text{fin,1}}.
 \label{IvrDfinal}
\end{eqnarray}

It is important to note that Eq.~(\ref{IvrDfinal}) was not
evaluated in the chirally symmetric phase; i.e., it can also be used
for studying the system before the chiral phase transition or in the
physical limit.  In the physical case we have an additional gap
equation for the mass $m$ that can be evaluated in the MSS procedure
by using the same manipulations used to evaluate $I_{\Delta}$. The $m
\to 0$ limit to study the chiral phase is trivial and it is taken in
the calculations  to be presented below.

%%%%%%%%%%%%%%%%%%%%%%%%%%%%%%%%%%%%%%%%%%%%%%%%%%%%%%%%%%%%%%%%%%%%
\section{Contrasting the TRS and MSS regularization procedures}
\label{results}

To obtain the numerical results for $N_{c}=3$ in $\beta$ equilibrium
we first evaluate the gap equation for $\Delta$, the charge neutrality
conditions for $\mu_{8}$ and $\mu_{e}$, and the densities from
Eq.~(\ref{OmegaT0m0}).  {}For comparison purposes, we will present the
results for both schemes, TRS and MSS.

%%%%%%%%%%%%%%%%%%%%%%%%%%%%%%%%%%%%%%%%%%%%%%%%%%%%%%%%%%%%%%%%%%%%%%%%%
\subsection{The $\Delta$ gap equation} 

The gap equation is given by 
\begin{eqnarray}
\frac{\partial\Omega_{T=0}}{\partial\Delta}\Bigr|_{\Delta=\Delta_c}
=0,
\end{eqnarray}
where $\Delta_c$ is the solution of
\begin{eqnarray}
1=2G_{d}\left[4I_{\Delta}^{i}
  -2\theta\left(\delta\mu-\Delta_c\right)\int_{\mu^{-}}^{\mu^{+}}\frac{dp}{\pi^{2}}
  \frac{p^{2}}{E_{\Delta_c}^{-}}\right],
\label{GapD}
\end{eqnarray}
where $I_{\Delta}$ is given, in the TRS and MSS cases respectively by
\begin{eqnarray}
I_{\Delta}^{TRS} & = &
\int_{\Lambda}\frac{d^{3}p}{\left(2\pi\right)^{3}}
\left(\frac{1}{E_{\Delta}^{+}}+\frac{1}{E_{\Delta}^{-}}\right),
\label{IdTRS}
\\ I_{\Delta}^{MSS} & = &
2I_{\text{quad}}-\left(\Delta^{2}-M_{0}^{2}-2\mu^{2}\right)
I{}_{\text{log}} + 2I_{\text{fin,2}}\nonumber \\ & + &
\left[\frac{3\left(M_{0}^{2}-\bar{\mu}^{2}-\Delta^{2}\right)^{2}}{4}
  -3M_{0}^{2}\bar{\mu}^{2}\right]I_{\text{fin,1}},
 \label{IdMSS}
\end{eqnarray}
where the indicated integrals were defined in the previous section,
Eqs.~(\ref{quad}) - (\ref{Ifin2}), in
the limit $m\to 0$, and $E_{\Delta}^{\pm}$ are  defined in
Eq.~(\ref{Edelta}).

%%%%%%%%%%%%%%%%%%%%%%%%%%%%%%%%%%%%%%%%%%%%%%%%%%%%%%%%%%%%%%%%%%%%%%%%%
\subsection{The color neutrality condition} 

It has been widely discussed in the literature that the
superconducting quark matter that may occur in compact stars is
required to be both electromagnetic and color neutral,  such as to be
in the stable bulk phase~\cite{Amore2002,Steiner2002,Huang2003}. The
color neutrality represents the equality between the numbers of quarks
with colors red, green, and blue,  since the quark matter must be
composed by color singlets.  In Ref.~\cite{Amore2002} the authors
have shown that once a macroscopic chunk  of color superconductor is
color neutral, implementation of the projection which  imposes color
singletness has a negligible effect on the free energy of the state,
similar to the usual fact from ordinary superconductivity. In this
case, the projection which turns a BCS state, wherein the particle number
is formally indefinite, into a state with definite but very large
particle number has no  significant effect.  Color singletness follows
without paying any further free energy price~\cite{Alford2002}.  It is
important to mention that by imposing these constraints, the free
energy of the 2SC phase becomes extremely large and cannot be found in
nature.  This problem disappears if we also consider the $s$ quark, in
which case the phase of the system  is CFL, which satisfies the
neutrality constraints, costing a smaller quantity of free energy. 

In this work we are focused on the correct separation of vacuum
divergences, from  finite integrals, as well as the influence of 
the regularization scheme on the values of the diquark condensate and, 
consequently, in the phase diagrams of the system when $\beta$ 
equilibrium and  charge neutrality are taken into account. 
{}For this reason, we will be working  simply
with the $SU(2)$ version of the NJL model, in which case, due to the
definitions~(\ref{muur}) to~(\ref{mudb}),  only quarks with red and
green colors do participate in the pairing.

The color neutrality condition is obtained by imposing that the number
density $n_{8}$ be vanishing.  Thus,  it is required that
\begin{eqnarray}
n_{8} & = & -\frac{\partial\Omega_{T=0}}{\partial\mu_{8}}=0,
\end{eqnarray}
which leads to the condition
\begin{eqnarray}
0 & = & -\frac{\mu_{ub}^{3}}{3\pi^{2}}-\frac{\mu_{db}^{3}}{3\pi^{2}}
+2I_{8}^{i}+2\bar{\mu}I_{\Delta}^{i} \nonumber \\ &+&
\theta\left(\delta\mu-\Delta\right)\int_{\mu^{-}}^{\mu^{+}}\frac{dp}{\pi^{2}}p^{2}
\, \frac{p-\bar{\mu}}{E_{\Delta}^{-}},
\label{n8}
\end{eqnarray}
with $I_{\Delta}^{i}$ defined in Eq. (\ref{IdTRS}) or (\ref{IdMSS}),
for the TRS or MSS cases,  respectively, and $I_{8}^{i}$ is given by 
\begin{eqnarray}
I_{8}^{TRS} & = & \int_{\Lambda}\frac{d^{3}p}{\left(2\pi\right)^{3}}
\left(\frac{p}{E_{\Delta}^{+}}-\frac{p}{E_{\Delta}^{-}}\right),
\label{I8TRS}
\\ I_{8}^{MSS} & = & -2\bar{\mu}I_{\text{quad}}
+\bar{\mu}\left(3\Delta^{2}-M_{0}^{2}-2\bar{\mu}^{2}\right)
I_{\text{log}} +I_{\text{fin,5}} \nonumber \\ & + &
M_{0}^{2}\bar{\mu}\left(3M_{0}^{2}+2\bar{\mu}^{2}-3\Delta^{2}\right)
I_{\text{fin,1}}
\nonumber \\ & + & \frac{5\bar{\mu}}{4}\left[4M_{0}^{2}\bar{\mu}^{2}
  -3\left(M_{0}^{2}-\bar{\mu}^{2}-\Delta^{2}\right)^{2}\right]
I_{\text{fin,4}} .
 \label{I8MSS}
\end{eqnarray}
In Appendix~\ref{mssEqs} we give some of the details on the explicit
calculation of $I_{8}^{MSS}$ and present there also the definitions
of $I_{\text{fin,4}}$ and $I_{\text{fin,5}}$.

%%%%%%%%%%%%%%%%%%%%%%%%%%%%%%%%%%%%%%%%%%%%%%%%%%%%%%%%%%%%%%%%%%%%%%%%%
\subsection{The electric neutrality condition} 

When considering the electric neutrality condition, we must note that
the integrands have the same structure as the ones in $n_{8}$.  Then,
by imposing that
\begin{eqnarray}
n_{e} & = & -\frac{\partial\Omega_{T=0}}{\partial\mu_{e}}=0,
\end{eqnarray}
we obtain that
\begin{eqnarray}
0 & = & \frac{\mu_{e}^{3}}{3\pi^{2}}-\frac{2\mu_{ub}^{3}}{9\pi^{2}} +
\frac{\mu_{db}^{3}}{9\pi^{2}} -\frac{2}{3}\bar{\mu}I_{\Delta}^{i}
-\frac{2}{3}I_{8}^{i} \nonumber \\ &+&
2\theta\left(\delta\mu-\Delta\right)\int_{\mu^{-}}^{\mu^{+}}
\frac{dp\,p^{2}}{\pi^{2}}
\left(\frac{1}{2}-\frac{p-\bar{\mu}}{6E_{\Delta}^{-}}\right),
\label{ne}
\end{eqnarray}
where  $I_{\Delta}^{i}$ and $I_{8}^{i}$ are given by Eqs.~(\ref{IdTRS})
and  (\ref{I8TRS}), for the TRS case, or by Eqs.~(\ref{IdMSS}) and
(\ref{I8MSS}), for the MSS case.

%%%%%%%%%%%%%%%%%%%%%%%%%%%%%%%%%%%%%%%%%%%%%%%%%%%%%%%%%%%%%%%
\subsection{The results}

%%%%%%%%%%%%%%%%%FIGURE01%%%%%%%%%%%%%%%%%%%
\begin{center}
\begin{figure}[!htb]
\subfigure[] {\includegraphics[width=7cm]{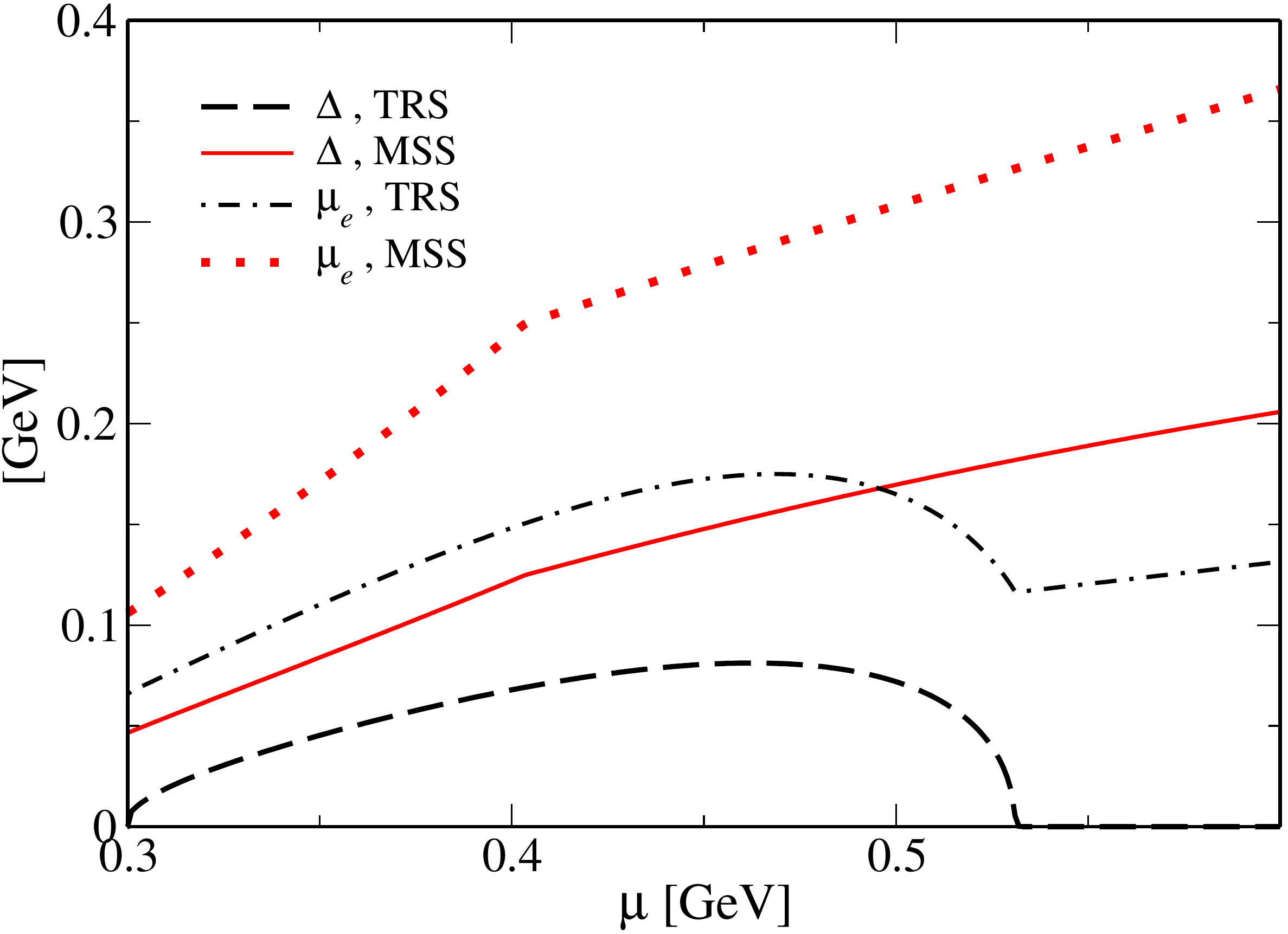} } \subfigure[]
          {\includegraphics[width=7cm]{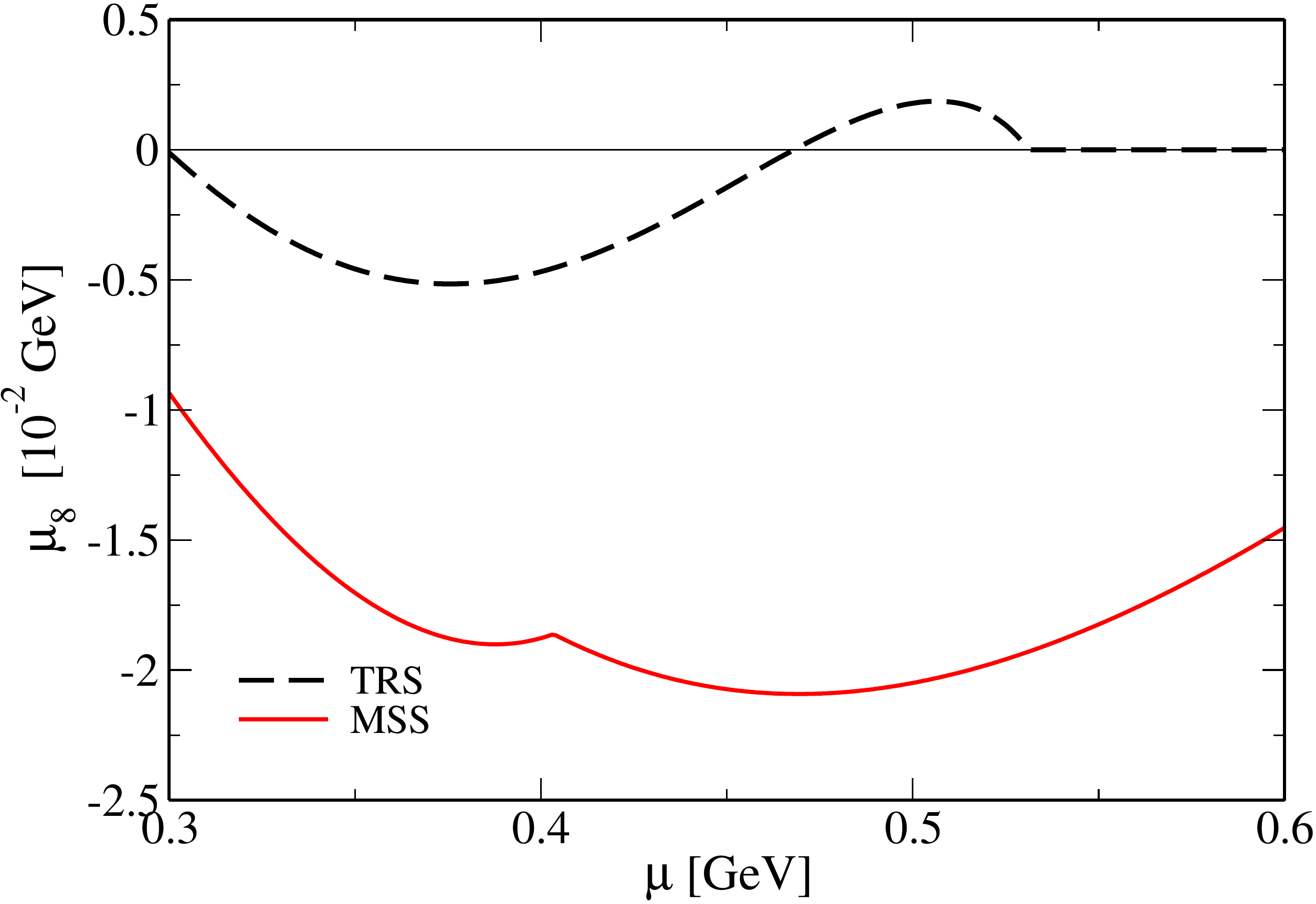}}
\caption{The diquark condensate $\Delta$ and the electron chemical
  potential $\mu_{e}$ (a) and the color chemical potential
  $\mu_{8}$ (b), as functions of $\mu$, for both the TRS and MSS
  regularization schemes. }
\label{D-mue-mu8}
\end{figure}
\end{center}
%%%%%%%%%%%%%%%%%%%%%%%%%%%%%%%%%%%%%%%%

To obtain our numerical results, we consider the values for the
parameters in the NJL model as chosen in the usual way, by the fitting
with the experimental vacuum values of the pion decay constant
$f_{\pi}=93$ MeV and chiral condensate $\left\langle
\bar{\psi}_{f}\psi_{f}\right\rangle ^{1/3}=-250$ MeV. The third
parameter, the use of the pion mass $m_{\pi}$, is not required if we
stay in the chiral limit.  The model parameters in this case are found
simply to be given by the values $G_{s}=5.0163$ GeV$^{-2}$ for the
scalar quark four-fermion interaction term, while the ultraviolet
cutoff scale is found to be $\Lambda=653.3$ MeV.  The diquark
coupling constant $G_{d}$ is set to be proportional to $G_{s}$, with
the value chosen as $G_{d}=0.75G_{s}$~\cite{BUBA,Huang2003}.

In {}Fig.~\ref{D-mue-mu8} we show the results for $\Delta,\, \mu_{e}$,
and $\mu_{8}$, obtained by solving numerically Eqs.~(\ref{GapD}),
(\ref{n8}), and (\ref{ne}) in both the TRS and MSS regularization
procedures.  Note that the result for $\Delta$ initially increases
with the chemical potential in both TRS and MSS regularization
procedures, but as $\mu$ gets sufficiently large, $\mu \sim 0.53$ GeV,
the result for TRS drops down and it is vanishing from then on. The
result for MSS still grows with $\mu$ as expected in general grounds.
Note also the change in behavior for both the chemical potentials
$\mu_e$ and $\mu_8$ in both regularization schemes. Even for small
values for the chemical potential, there are significant quantitative
differences in the results obtained in the TRS and MSS procedures.

\begin{center}
\begin{figure}[!htb]
\subfigure[] {\includegraphics[width=7cm]{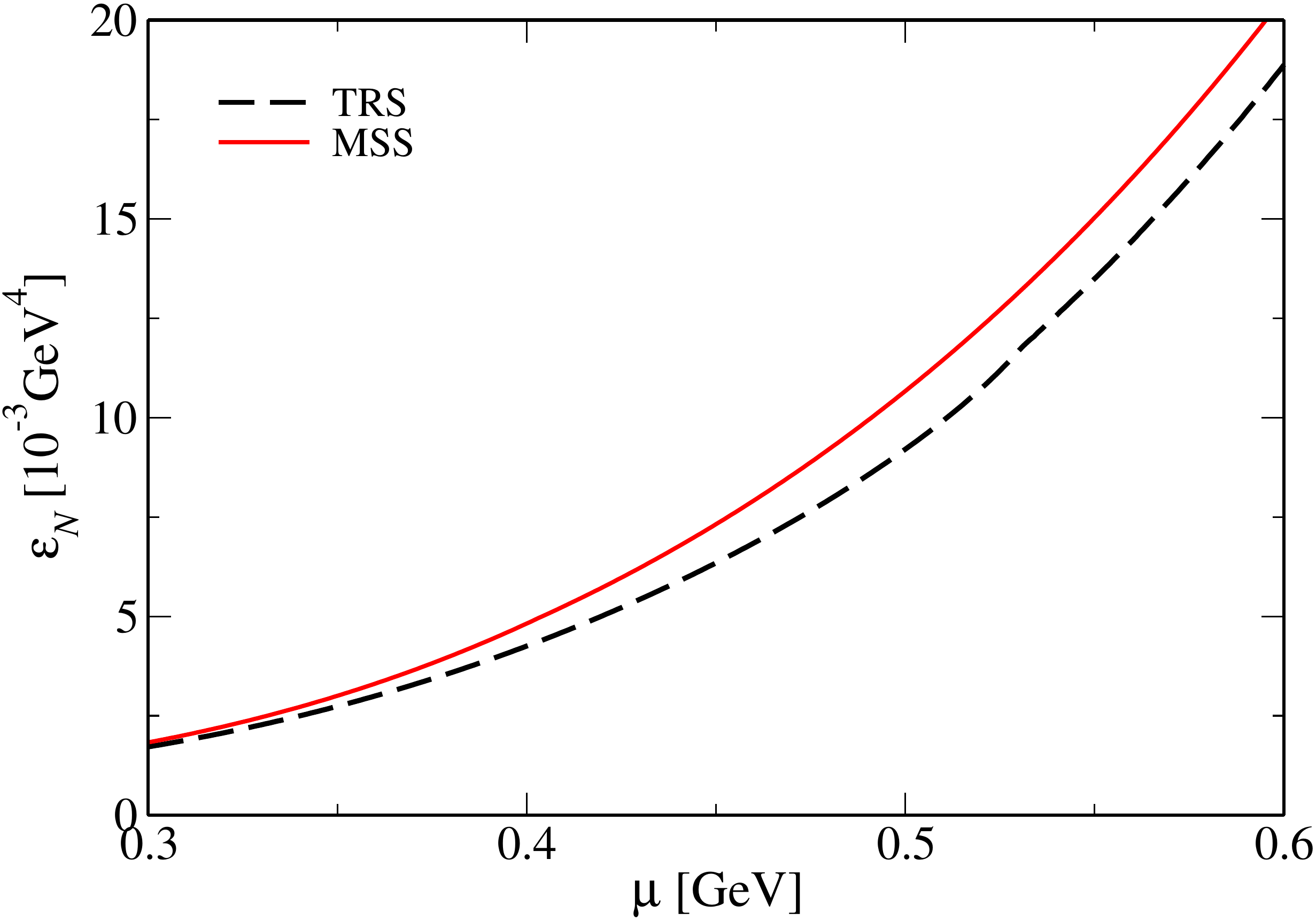} } \subfigure[]
          {\includegraphics[width=7cm]{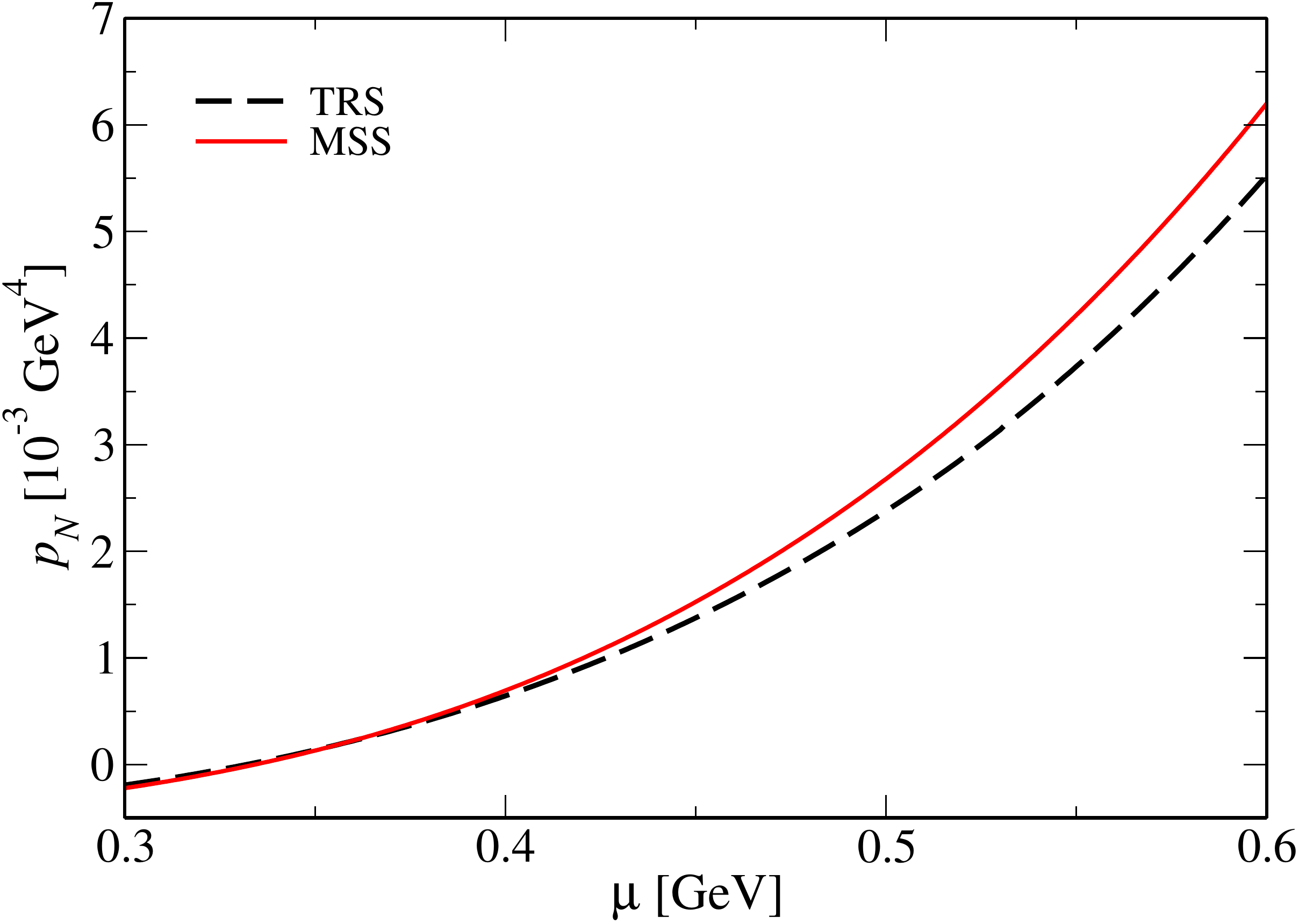}} \subfigure[]
          {\includegraphics[width=7cm]{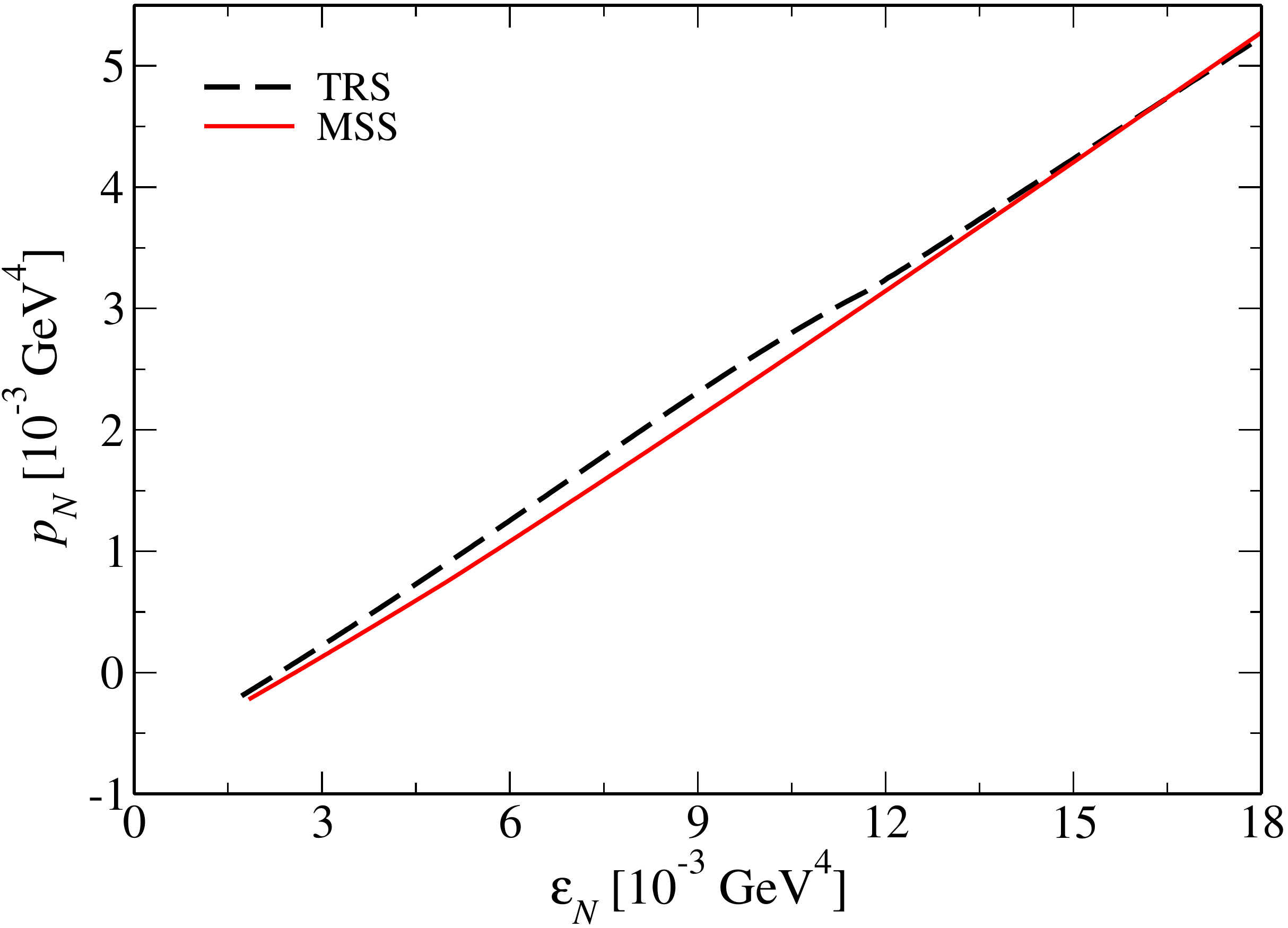}}
\caption{The energy density (a), the normalized pressure (b), and the
  equation of state (c) as functions of $\mu$,  for both the TRS and
  MSS regularization schemes. }
\label{thermo}
\end{figure}
\end{center}

\begin{center}
\begin{figure}[!htb]
\subfigure[] {\includegraphics[width=7cm]{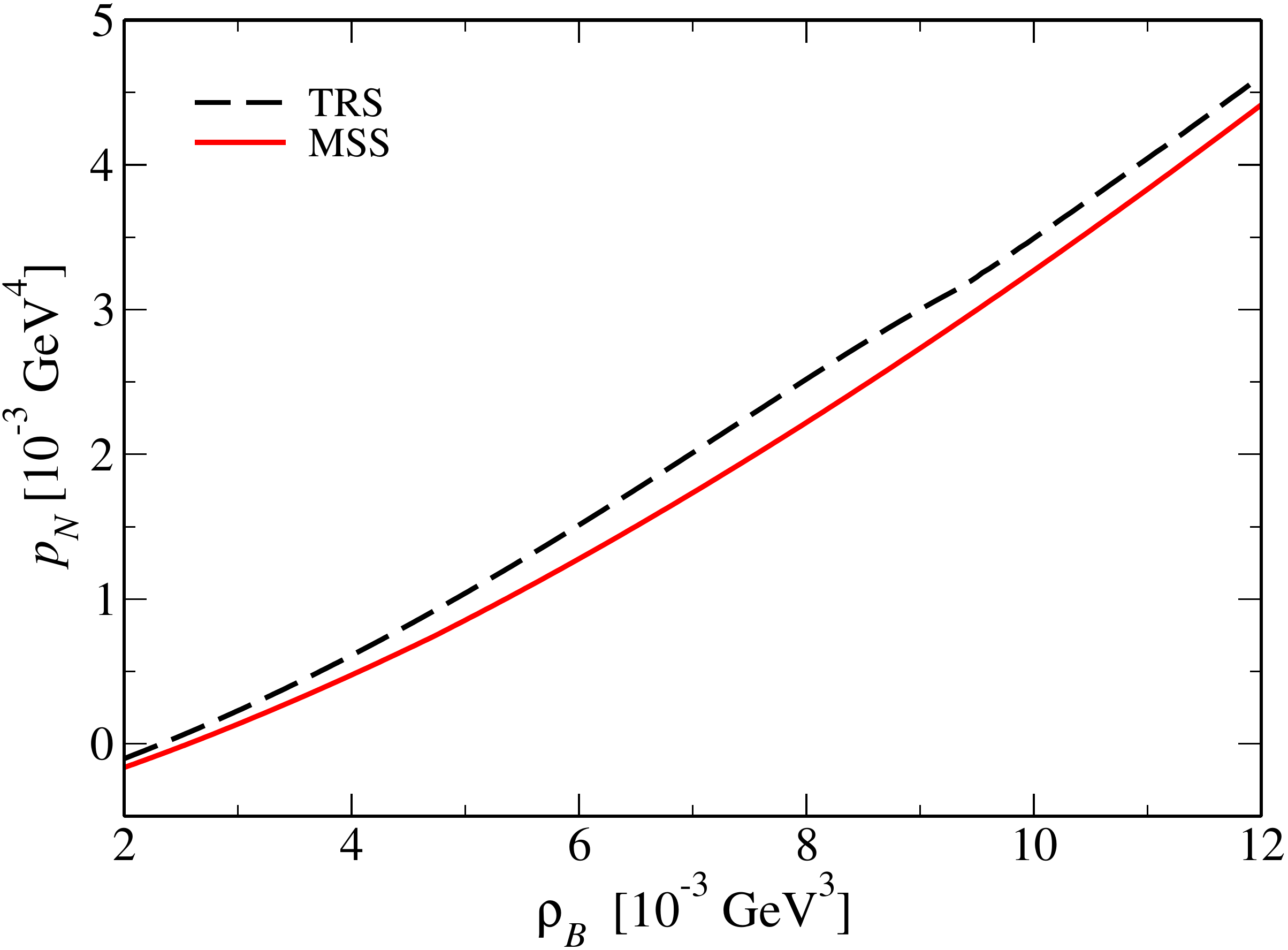} } \subfigure[]
          {\includegraphics[width=7cm]{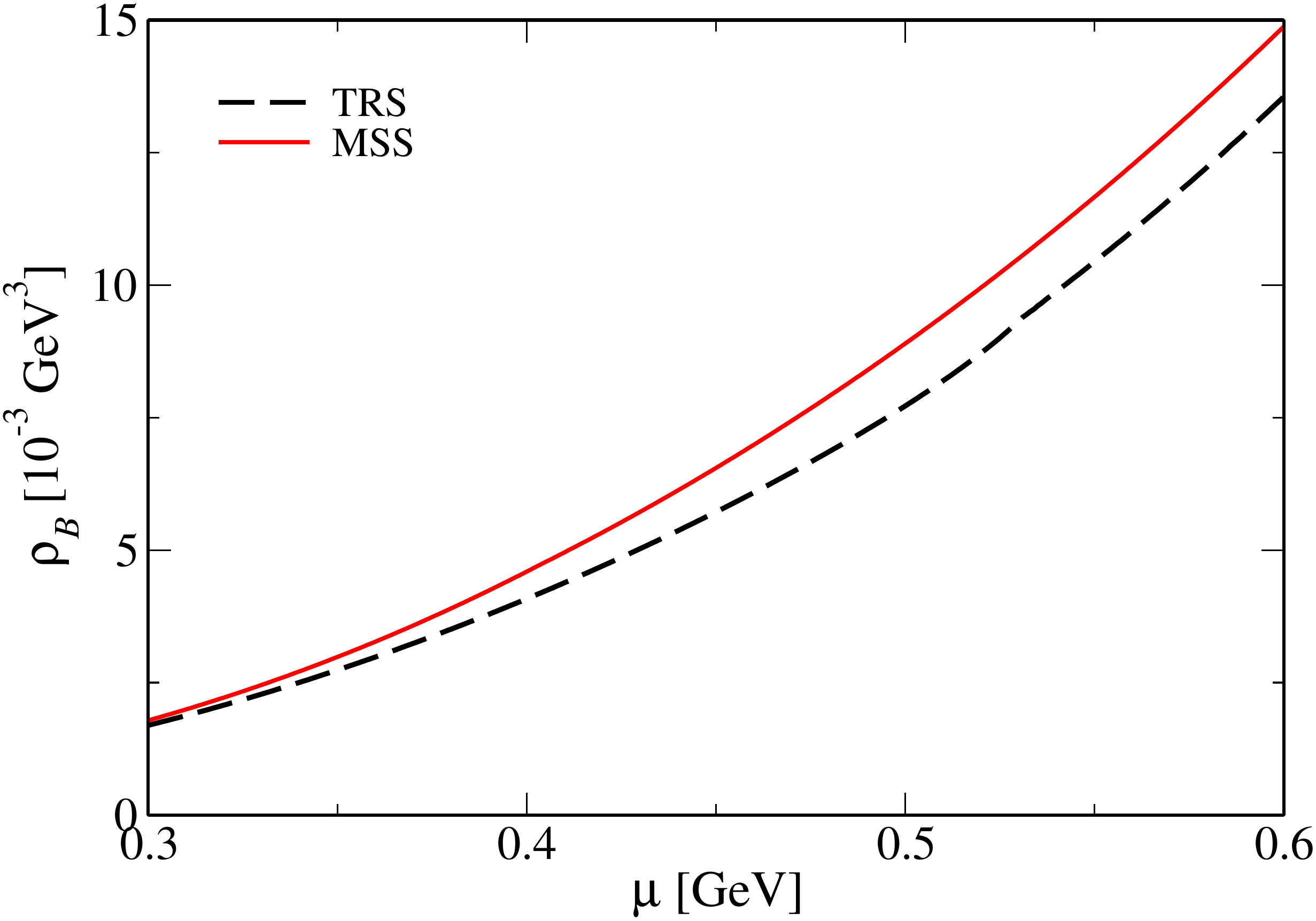}}
\caption{The normalized pressure as a function of the baryon density
  (a) and the baryon and individual densities $\rho_{u}$ and
  $\rho_{d}$ (b) as functions of the chemical potential $\mu$, for
  both the TRS and MSS regularization schemes. }
\label{dens}
\end{figure}
\end{center}

To emphasize the differences between the TRS and MSS, we also show some
of the relevant thermodynamic quantities to compare them between the
two methods. {}First of all, to obtain the baryon density $\rho_{B}$
we need to determine the total density $\rho_{T}=\rho_{u}+\rho_{d}$ in
the $SU(2)$ case. However, our expression for the thermodynamic
potential, Eq.~(\ref{OmegaT0m0}), is written in terms of $\bar{\mu}$
and $\delta\mu$.  Here, we restrict ourselves to show the final
expressions for each quantity, whose details for their derivation are
given in the Appendix~\ref{CalcDensities}.  The individual densities
$\rho_{u}$ and $\rho_{d}$ are given by 
\begin{eqnarray}
\rho_{u} & = & \frac{\mu_{ub}^{3}}{3\pi^{2}} + 2I_{8}^{i}
+2\bar{\mu}I_{\Delta}^{i}+\theta\left(\delta\mu-\Delta\right)
\int_{\mu^{-}}^{\mu^{+}}\frac{dp}{\pi^{2}}p^{2}
\left(\frac{p-\bar{\mu}}{E_{\Delta}^{-}}\right)
\nonumber \\ & - & \frac{2\sqrt{\delta\mu^{2}-\Delta^{2}}}{3\pi^{2}}
\left(\delta\mu^{2}-\Delta^{2}+3\bar{\mu}^{2}\right)
\theta\left(\delta\mu-\Delta\right),
 \label{rhou}
\end{eqnarray}
and
\begin{eqnarray}
\rho_{d} & = & \frac{\mu_{db}^{3}}{3\pi^{2}} +
2I_{8}^{i}+2\bar{\mu}I_{\Delta}^{i}+
\theta\left(\delta\mu-\Delta\right)
\int_{\mu^{-}}^{\mu^{+}}\frac{dp}{\pi^{2}}p^{2}
\left(\frac{p-\bar{\mu}}{E_{\Delta}^{-}}\right)
\nonumber \\ & + & \frac{2\sqrt{\delta\mu^{2}-\Delta^{2}}}{3\pi^{2}}
\left(\delta\mu^{2}-\Delta^{2}+3\bar{\mu}^{2}\right)
\theta\left(\delta\mu-\Delta\right),
 \label{rhod}
\end{eqnarray}
where $I_{8}^{i}$ and $I_{\Delta}^{i}$ were already defined previously
for each of the two schemes.  The normalized pressure
$p_{N}$, the baryon density $\rho_{B}$ and the energy density
$\varepsilon$ are then given, respectively, by
\begin{eqnarray}
p_{N} & = & -\Omega_{T=0}\left(\Delta,\bar{\mu},\delta\mu\right),
\label{pressN}\\
\rho_{B} & = & \frac{\rho_{T}}{3},
\label{rhobaryon}\\
\varepsilon & = &
-p_{N}+\mu_{B}\rho_{B}\left(=-p_{N}+3\mu\rho_{B}\right).
\label{epsilon}
\end{eqnarray}
The numerical results for these quantities, as well as the individual
densities and the equation of state, $p_{N}\times\varepsilon_{N}$, are
shown in  {}Figs.~\ref{thermo} and~\ref{dens}.  Though the
differences in the EoS are not very significant in both the
regularization schemes, the energy density and pressure tend to
increase faster in the MSS case than in the TRS one. 

%%%%%%%%%%%%%%%%%FIGURE02%%%%%%%%%%%%%%%%%%%
%%%%%%%%%%%%%%%%%%%%%%%%%%%%%%%%%%%%%%%%

%%%%%%%%%%%%%%%%%FIGURE03%%%%%%%%%%%%%%%%%%%
%%%%%%%%%%%%%%%%%%%%%%%%%%%%%%%%%%%%%%%%

%%%%%%%%%%%%%%%%%FIGURE04%%%%%%%%%%%%%%%%%%%
\begin{center}
\begin{figure}[!htb]
\includegraphics[width=7cm]{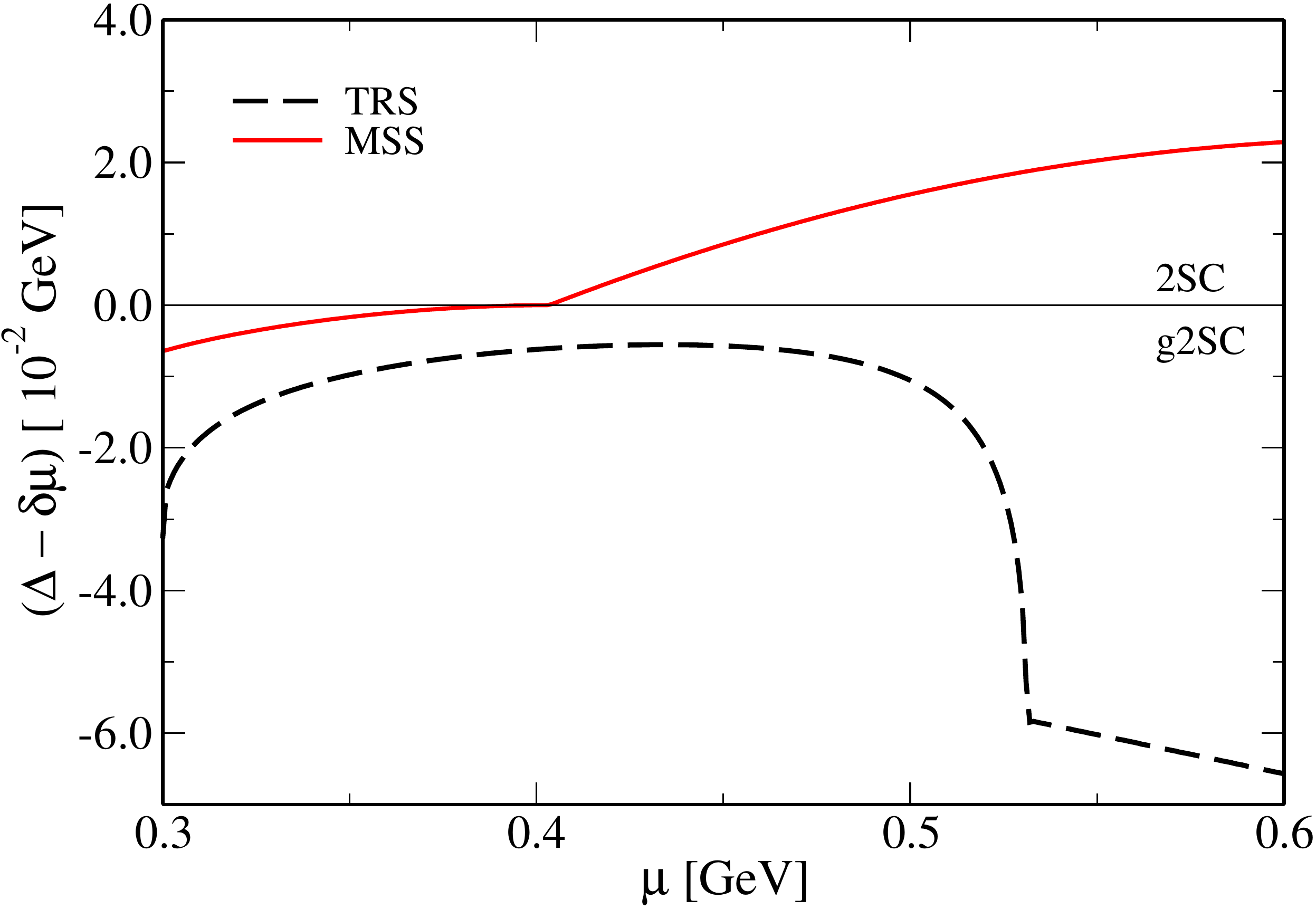} 
\caption{The criterion $\delta\mu-\Delta$ used to define a g2SC or 2SC
  phase of the system in the cases of the TRS and MSS regularization
  procedures.}
\label{g2sc}
\end{figure}
\end{center}
%%%%%%%%%%%%%%%%%%%%%%%%%%%%%%%%%%%%%%%%

It is argued in the literature~\cite{Huang:2003xd,Shovkovy:2003uu}
that a neutral 2SC phase exists at low values of $\mu$, called
th ``gapless-2SC'' (g2SC) phase, instead of the usual 2SC one (for a
detailed discussion regarding the gapless  phase, structure, and
consequences, see in addition~\cite{revCSC} and references therein).
The criterion used to define whether the system is in the gapless or
in the 2SC phase is quite simple. In $\beta$ equilibrium the new
contribution to the effective potential is exactly the term that
contains a step function on Eq.~(\ref{OmegaT0m0}). If
$\delta\mu>\Delta$, this term remains and the phase is said to be in
the g2SC. Otherwise, for $\delta\mu<\Delta$, that term disappears and
the phase is said to be in the usual 2SC one. In {}Fig.~\ref{g2sc} we
probe the emergence of these two possible phases.

%%%%%%%%%%%%%%%%%FIGURE5%%%%%%%%%%%%%%%%%%%
\begin{center}
\begin{figure}[!htb]
\subfigure[] {\includegraphics[width=7cm]{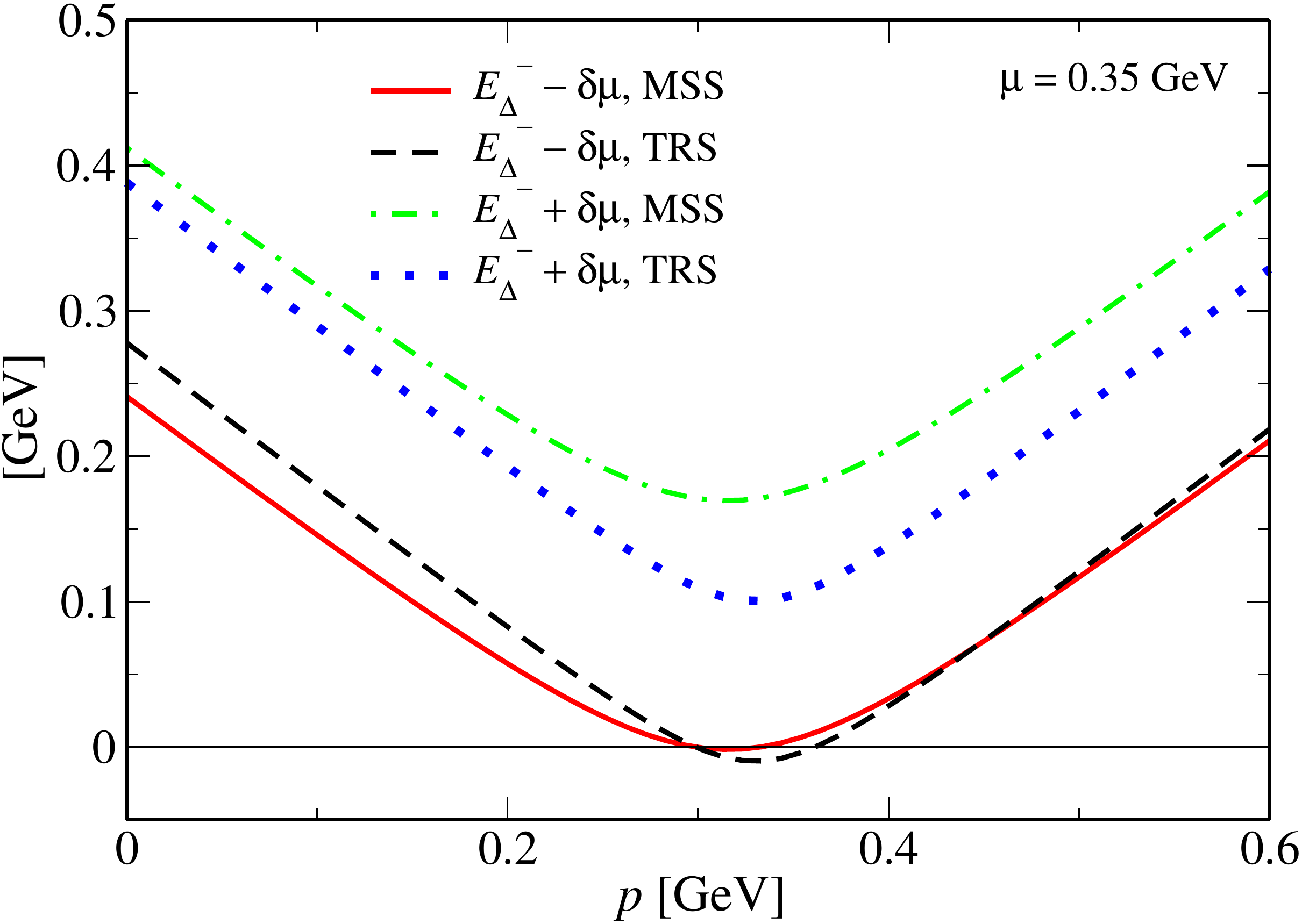} } \subfigure[]
          {\includegraphics[width=7cm]{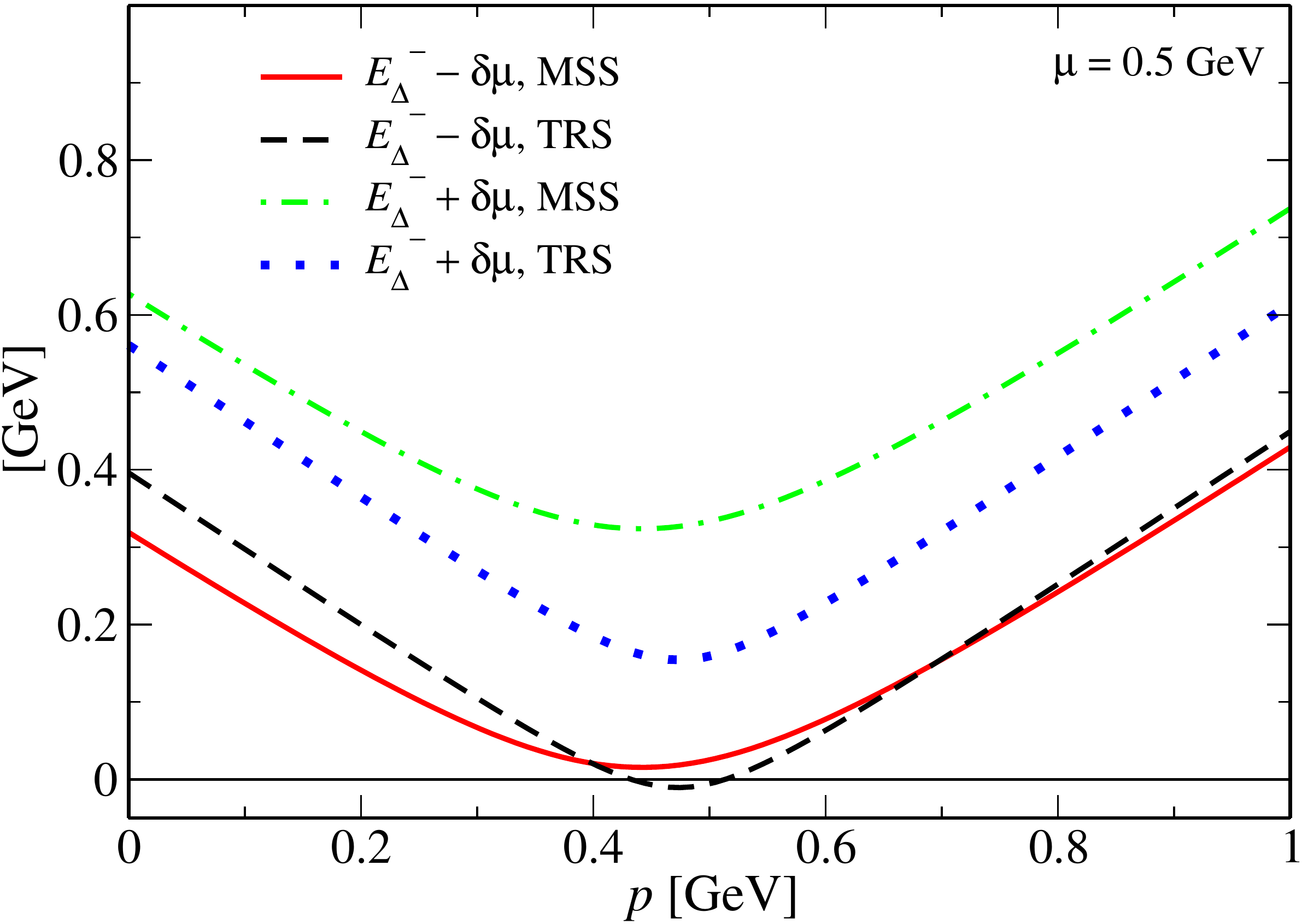}}
\caption{Quasiparticle dispersion relations $E_{\Delta}^{-}\pm
  \delta\mu$, for  (a) $\mu = 0.35$ GeV and (b) $\mu = 0.5$ GeV.}
\label{dispersions}
\end{figure}
\end{center}
%%%%%%%%%%%%%%%%%%%%%%%%%%%%%%%%%%%%%%%%

It is also useful to discuss the difference between the
g2SC and 2SC phases from the dispersion  relations for
quasiparticles in the context of the two regularization schemes
studied here.  {}From Eqs.~(\ref{rhou}) and~(\ref{rhod}) one can see
that  in g2SC the pairing quarks have different number densities,
which does not occur in 2SC.  The spectrum in the 2SC case includes
the free blue quark that does not take place in the pairing and also
the quasiparticle excitations formed by the linear superposition of
$u_{rg}$ and  $d_{rg}$ quarks,  whose gap is $\Delta$. 
When $\delta\mu \neq \Delta$, there is a small  discrepancy between the
{}Fermi surfaces of the $u$ and $d$ pairing quarks, shifting  one of
the dispersions to $\Delta + \delta\mu$ and the other to $\Delta -
\delta\mu$.  When the mismatch $\delta\mu > \Delta$, the lower
dispersion relation becomes  negative for some values of $p$ and this
is the spectrum usually called gapless.  This is illustrated in
{}Fig.~\ref{dispersions}. {}For $\mu = 0.35$ GeV, both regularization
schemes  predict a gapless phase for the system, even though in the
MSS case the value of $E_{\Delta}^- - \delta\mu$  is smaller (in
magnitude) than in the TRS case. On the other hand, for $\mu = 0.5$
GeV, the dispersion for  the MSS never becomes negative; i.e., the
system is in the 2SC phase, while in the TRS the  gapless phase is
observed again.

One can see that using the MSS regularization procedure the behavior
of the phase structure can become quite different. While in the TRS
case the system is always in the g2SC phase in the range of $\mu$
considered in this work, the situation becomes quite the opposite in
the case of the MSS regularization procedure. We can see from the
results shown in {}Figs.~\ref{g2sc}
and~\ref{dispersions} that now the system can display
a transition between the g2SC and the 2SC phases.  This is quite a
remarkable difference and it is the main result of this work. We can
trace this change of behavior in the MSS case by recalling the result
shown in {}Fig.~\ref{D-mue-mu8} for the diquark condensate
$\Delta$. In the TRS case, the diquark condensate never increases to be
above $\delta \mu$ and, even worse, it vanishes after $\mu \gtrsim
0.53$ GeV. However, in the MSS, $\Delta$ is larger than in the TRS
case and always increases with the chemical potential.  Thus, it is no
wonder that at some point it will become larger than $\delta \mu$ and
the system can transit from a g2SC to a 2SC phase. It is nice to see
that this can happen already for not relatively too large values of
the chemical potential. 
This is also reflected on the behavior of
thermodynamic quantities. {}From {}Figs.~\ref{thermo}
and~\ref{dens}, one notices that the difference between two schemes
increases with the increase of  the chemical potential. This is
related to the change to the 2SC phase in the MSS case, while in the
TRS case the system is still in the g2SC phase.

\begin{center}
\begin{figure}[!htb]
\subfigure[]{\includegraphics[width=7cm]{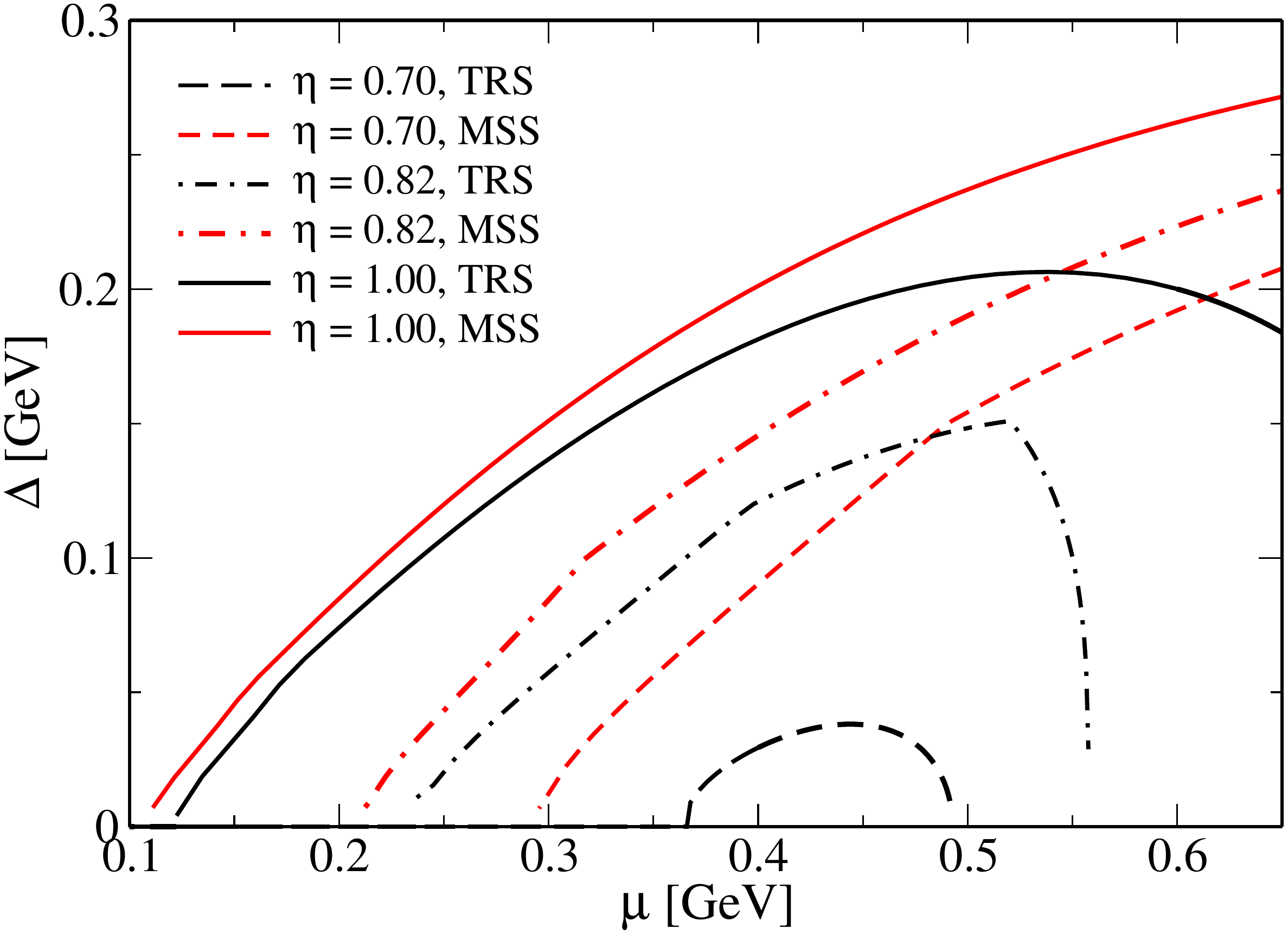} } 
\subfigure[]{\includegraphics[width=7cm]{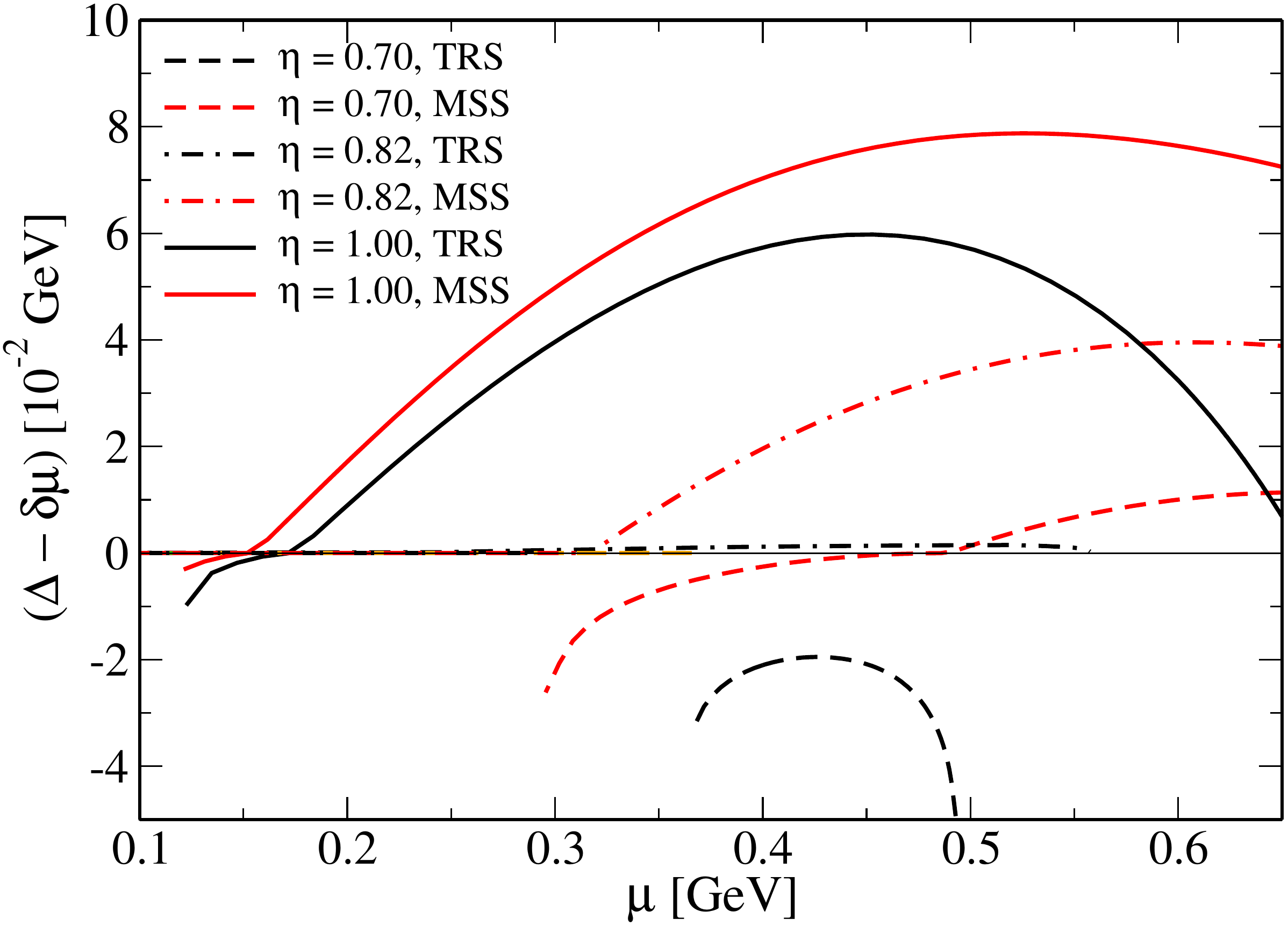}}
\caption{Diquark condensate  (a) and the change from the g2SC to the 2SC 
phases (b), as a function of the chemical potential and for different 
values of the ratio  $G_d/G_s\equiv \eta$.}
\label{Gdfree}
\end{figure}
\end{center}

{}Finally, it is also useful to comment on the effect of the value of the
diquark interaction $G_d$ on the results. In all of the above results, we have used
$G_d=3G_s/4$, a value naturally motivated when deriving the Lagrangian density 
Eq.~(\ref{lag}) from the QCD one-gluon exchange approximation and it results
from the {}Fierz transform of the primary color current-current coupling, 
projected into the relevant quark-antiquark and diquark channels~\cite{BUBA}.
However, it is quite common in the literature to simply consider $G_d$ as an
additional free parameter of the model. In {}Fig.~\ref{Gdfree}(a) we show the
results for the diquark condensate and in {}Fig.~\ref{Gdfree}(b) the results
equivalent to {}Fig.~\ref{g2sc}, as a function of some representative values of
the ratio $G_d/G_s\equiv \eta$, to illustrate the differences that appear 
when using the TRS or the MSS procedures.
%
%%%%%%%%%%%%%%%%%FIGURE6%%%%%%%%%%%%%%%%%%%
%%%%%%%%%%%%%%%%%%%%%%%%%%%%%%%%%%%%%%%%
%
{}From the results shown in {}Fig.~\ref{Gdfree}, we see that the differences between 
the TRS and MSS procedures are more pronounced for values of $\eta <1$. As we increase
$\eta$, the differences decrease, but they are still evident and quantitatively large
as the chemical potential increases. In particular, we still see a decrease of the
diquark condensate $\Delta$ in the TRS case even when $\eta=1$. {}For a ratio
of $\eta \gtrsim 0.82$, we see from {}Fig.~\ref{Gdfree}(b) that the TRS can also
display a transition from the g2SC to the 2SC phase, yet, the difference with the
MSS case is always appreciable. 
It is also important to mention that if we consider lower values of $\eta$, the 
system is in the gapless phase for larger values of $\mu$ in MSS, and the value of 
$\Delta$ becomes smaller for both schemes.
% Particularly, for $\eta = 0.7$ MSS still predicts the 2SC phase, 
% as we may see in {}Fig.~\ref{Gdfree}.
For $\eta \sim 0.68$, we did not find solutions for $\Delta \neq 0$ using TRS, and the 
color superconducting phase is not predicted in this scheme.

%%%%%%%%%%%%%%%%%%%%%%%%%%%%%%%%%%%%%%%%%%%%%%%%%%%%%%%%%%%%%%%%%%%%%%%%%%%%%
\section{Conclusions}
\label{conclusions}

In this work we have studied an alternative regularization approach
where medium effects are explicitly separated from vacuum dependent
terms and UV divergent momentum integrals become only dependent on the
vacuum quantities.  We have applied this to the NJL model with diquark
interactions and in $\beta$ equilibrium.  We perform an explicit
comparison of this proposed scheme, called the MSS regularization
procedure, with the more traditional one,  the TRS regularization
procedure, where no separation of vacuum and medium effects is
done. Our results point to both qualitatively and quantitatively important
differences between these two methods, in particular regarding the phase
structure of the model in the cold and dense nuclear matter
case. While in the TRS case the diquark condensate eventually vanishes
for a chemical potential of order $\mu\sim 0.53$ GeV, in the MSS case
the diquark condensate always increases.  As a consequence of this
result, we show that there is a change in the phase structure of the
model from a g2SC to a 2SC phase that is also reflected on the thermodynamic
quantities.
The phase structure, is also affected by the value of the ratio 
$G_d/G_s$, since a large value of the diquark constants favors the 2SC over the 
g2SC.

{}Finally, given that the differences between the MSS
and TRS approaches becomes more pronounced at  high densities, it is
important also to comment on the validity of the two-flavor
approximation used here.  It is known that the charge neutrality has
a strong influence in the effective $s$ quark mass, which starts to
decrease at around  $\mu \sim 400$ MeV, in comparison with the case
without neutrality~\cite{Steiner:2002gx}. Near the {}Fermi surface
($p\sim\mu$)  the dispersion can be approximated as $\sqrt{p^2 +
  M_s^2} - \mu \sim p - \left(\mu - \frac{M_s^2}{2\mu}\right)$.
Since $M_s \simeq$ 150 MeV is a good estimate for the $s$ quark
mass in the intermediate density region, one notices that the
superconductor gap $\Delta$ has the same order of the scale
$\frac{M_s^2}{2\mu}$. {}For this reason, the contribution effects
due to the $s$ quark should be  taken into account and keeping in
mind that  the free energy cost to satisfy the neutrality conditions
is too large in 2SC, the CFL  phase is favored~\cite{Alford2002}. In
either case, as the driving difference between the TRS and MSS
approaches is in the way the medium effects are handled, it is quite
reasonable to expect that even in the case of including the effects
of the $s$ quark, the differences in the results will remain. These
differences might even increase as additional density effects due to
the $s$ quark are added and also in the case of including thermal 
and external magnetic fields effects~\cite{Abhishek:2018xml}.
In future works it will be interesting to further analyze
these differences between the TRS and MSS regularization approaches,
as also when going beyond the mean-field
approximation~\cite{Duarte:2017zdz} which can possibly exacerbate
the differences between results derived when using either the TRS or
MSS approaches.

%%%%%%%%%%%%%%%%%%%%%%%%%%%%%%%%%%%%%%%%%%%%%%%%%%%%
\acknowledgments

This work was partially supported by  Funda\c{c}\~ao de Amparo \`a
Pesquisa do Estado de S\~ao Paulo (FAPESP) under Grant No. 2017/26111-4
(D. C. D.),  Conselho Nacional de Desenvolvimento Cient\'ifico e
Tecnol\'ogico (CNPq)  under Grants  No. 304758/2017-5 (R. L. S. F.) and
No. 302545/2017-4 (R. O. R.)  and Funda\c{c}\~ao Carlos Chagas Filho de Amparo
\`a Pesquisa do Estado do Rio de Janeiro (FAPERJ)  under Grant No.
E-26/202.892/2017 (R. O. R.).  D. C. D. and R. L. S. F. also thank N. N. Scoccola
and M. Coppola for useful discussions.

%%%%%%%%%%%%%%%%%%%%%%%%%%%%%%%%%%%%%%%%%%%%%%%%%%%%

\appendix

%%%%%%%%%%%%%%%%%%%%%%%%%%%%%%%%%%%%%%%%%%%%%%%%%%%%%%%%%%%%%%%%%%%%%%%%
\section{Color neutrality condition and thermodynamic potential in the 
MSS regularization scheme}
\label{mssEqs}

In Eq.~(\ref{OmegaT}) we have the full expression for the
thermodynamic potential at finite temperature and
$\beta$ equilibrium. With the exception of the last momentum integral in
there, which needs to be regularized, all other terms are finite. In
this appendix we show some details in the calculation of the
contributions that come from this integral when using the MSS
regularization scheme.  In particular, the expression of the color
neutrality condition, $n_{8}=0$, has an unusual usual divergency structure,
so the procedure to separate the divergent integrals requires some
extra care, which we here explain in detail. 
Once more we choose to present the calculations for
the more general case,  i. e., in the physical limit, since the
chiral limit is trivial to obtain from  the final expressions. In
this way, the dispersions needed here are defined in Eqs.~(\ref{E})
and (\ref{Ed}). We start from the correspondent integrand that comes
from Eq.~(\ref{OmegaT}), when we derive it with respect to $\mu_{8}$,
which is
\begin{eqnarray}
\lefteqn{ \int\frac{d^{3}p}{\left(2\pi\right)^{3}}
  \left(\frac{E+\bar{\mu}}{E_{\Delta}^{+}}-
  \frac{E-\bar{\mu}}{E_{\Delta}^{-}}\right)
} \nonumber \\ && =
I_{8}+\bar{\mu}\int\frac{d^{3}p}{\left(2\pi\right)^{3}}
\left(\frac{1}{E_{\Delta}^{+}}
+\frac{1}{E_{\Delta}^{-}}\right).
\label{intmu8}
\end{eqnarray}
Note that the second integrand in the right-hand side in the above
equation  is exactly $I_{\Delta}$ given by Eq.~(\ref{IvrDfinal}),
while $I_{8}$ is given by 
\begin{equation}
I_{8}^{i}=\int\frac{d^{3}p}{\left(2\pi\right)^{3}}\left(\frac{E}{E_{\Delta}^{+}}
-\frac{E}{E_{\Delta}^{-}}\right)
=\sum_{s=\pm1}\int\frac{d^{3}p}{(2\pi)^{3}} \frac{sE}{E_{\Delta}^{s}}.
\end{equation}
To deal with the UV divergence structure of $I_{8}$, we start by
writing it as 
\begin{eqnarray}
I_8 &=&
\sum_{s=\pm1}\int\frac{d^{3}p}{(2\pi)^{3}}\frac{sE}{E_{\Delta}^{s}}
\nonumber \\ & =&
\int\frac{d^{3}p}{(2\pi)^{3}}E\sum_{s=\pm1}s\left[\frac{1}{\pi}
  \int_{-\infty}^{+\infty}\frac{dp_{4}}{p_{4}^{2}+
    \left(E_{\Delta}^{s}\right)^{2}}\right].
\label{T1d4}
\end{eqnarray}
The procedure used here is quite similar to the one used in
Sec.~\ref{sec3}, with the difference that it is necessary to make one
more iteration of the identity~(\ref{ident}), to obtain 
\begin{eqnarray}
\lefteqn{ \frac{1}{p_{4}^{2}+\left(E+s\bar{\mu}\right)^{2}+\Delta^{2}}
  = }  \nonumber \\ && \frac{1}{p_{4}^{2}+p^{2}+M_{0}^{2}}
+\frac{A-2sE\bar{\mu}}{\left(p_{4}^{2}+p^{2}+M_{0}^{2}\right)^{2}}
\nonumber \\ & & +
\frac{\left(A-2sE\bar{\mu}\right)^{2}}{\left(p_{4}^{2}+p^{2}+M_{0}^{2}\right)^{3}}
+\frac{\left(A-2sE\bar{\mu}\right)^{3}}
{\left(p_{4}^{2}+p^{2}+M_{0}^{2}\right)^{4}}
\nonumber \\ & & +
\frac{\left(A-2sE\bar{\mu}\right)^{4}}{\left(p_{4}^{2}+p^{2}+M_{0}^{2}\right)^{4}
  \left[p_{4}^{2}+\left(E+s\bar{\mu}\right)^{2}+\Delta^{2}\right]}.
\label{r4it}
\end{eqnarray}
with $A=M_{0}^{2}-M^{2}-\bar{\mu}^{2}-\Delta^{2}$.  After some
algebraic manipulations and performing the sum over $s$ and the
$p_{4}$ integrations as indicated in Eq.~(\ref{T1d4}), we obtain 
\begin{eqnarray}
I_{8}^{MSS} & = & -2\bar{\mu}I_{\text{quad}}
+\bar{\mu}(2M_{0}^{2}-5\bar{\mu}^{2}-3A-2M^{2})I_{\text{log}}
\nonumber \\ & + &
\bar{\mu}(3M_{0}^{2}A+5M_{0}^{2}\bar{\mu}^{2}-3M^{2}A)I_{\text{fin,1}}
\nonumber \\ & -&
\frac{5m^{2}\bar{\mu}}{4}(3A^{2}+4M^{2}\bar{\mu}^{2})I_{\text{fin,3}}
\nonumber \\ & + &
\frac{5\bar{\mu}}{4}(4M_{0}^{2}\bar{\mu}^{2}-3A^{2}-8M^{2}\bar{\mu}^{2})
I_{\text{fin,4}}
+I_{\text{fin,5}}.  \nonumber \\
 \label{rI8}
\end{eqnarray}
where $I_{\rm quad}, \, I_{\rm log}, \, I_{\rm fin,1}$, and $I_{\rm
  fin,2}$ were already defined in Sec.~\ref{sec3}, while  $I_{\rm
  fin,3}, \, I_{\rm fin,4}$, and $I_{\rm fin,5}$ are given by
\begin{eqnarray}
I_{\text{fin,3}} & = & \int\frac{d^{3}p}{(2\pi)^{3}}
\frac{1}{\left(p^{2}+M_{0}^{2}\right)^{\frac{7}{2}}}, \nonumber
\\ I_{\text{fin,4}} & = & \int\frac{d^{3}p}{(2\pi)^{3}}
\frac{p^{2}}{\left(p^{2}+M_{0}^{2}\right)^{\frac{7}{2}}}, \nonumber
\\ I_{\text{fin,5}} & = &
\frac{35}{32}\sum_{s=\pm1}\int\frac{d^{3}p}{(2\pi)^{3}}
\int_{0}^{\infty}dt \frac{t^3}{\sqrt{1+t}} \nonumber \\ &\times&
\frac{sE(A-2sE\bar{\mu})^{4}t^{3}}{
  \left[(p^{2}+M_{0}^{2})t+(E+s\bar{\mu})^{2}+\Delta^{2}\right]^{\frac{9}{2}}},
\end{eqnarray}
where for $I_{\text{fin,5}}$ we have made use of the Feynman
parametrization formula,
\begin{equation}
\frac{1}{a^{m}b^{n}}=\frac{\Gamma(m+n)}{\Gamma(m)\Gamma(n)}\int_{0}^{\infty}
\frac{t^{m-1}dt}{(at+b)^{m+n}}.
\label{FeynParam}
\end{equation}
The only UV divergences now are in the first line of Eq.~(\ref{rI8})
above, but they do not have any dependence on medium terms. Note that
this expression is also used in the charge neutrality condition {[}see
  Eq.~(\ref{ne}){]}, as well as in Eq.~(\ref{IvrDfinal}).  To
obtain Eq.~(\ref{I8MSS}) used in Sec.~\ref{results}, we simply set
$M=0$ in the above equations.

%%%%%%%%%%%%%%%%%%%%%%%%%%%%%%%%%%%%%%%%%%%%%%%%%%%%%%%%%%%%%%%%%%%%%%
\section{The normalized thermodynamic
potential in the MSS regularization procedure}
\label{mssOmega}

To obtain the MSS expression for the normalized thermodynamic
potential, we use Eq.~(\ref{OmegaT}) to define
\begin{eqnarray}
\Omega_{N} & = & \Omega_{\text{finite}}+\Omega_{T}+\Omega_{\text{Reg}}
\nonumber \\ &=&-
\frac{\mu_{e}^{4}}{12\pi^{2}}+\frac{\left(M-m\right)^{2}}{4G_{s}}
-\frac{\left(M_{0}-m\right)^{2}}{4G_{s}}+\frac{\Delta^{2}}{4G_{d}}
\nonumber \\ & + &
\frac{T^{2}\mu_{e}^{2}}{6}+\frac{7\pi^{2}}{180}T^{4} \nonumber \\ &-&
\sum_{a}n_{a}\frac{2}{\beta}
\int\frac{d^{3}p}{\left(2\pi\right)^{3}}\; \ln\left(1+e^{-\beta
  E_{a}}\right) \nonumber \\ & + &
12\int\frac{d^{3}p}{\left(2\pi\right)^{3}}\sqrt{p^{2}+M_{0}^{2}}-\sum_{a}n_{a}
\int\frac{d^{3}p}{\left(2\pi\right)^{3}}E_{a}, \nonumber \\
\label{OmegaSep} 
\end{eqnarray}
where $\Omega_{\text{finite}}$  corresponds to pure mean field and
electron gas contributions; $\Omega_{T}$ is the temperature dependent
contributions; and $\Omega_{\text{Reg}}$ is the last two terms in
Eq.~(\ref{OmegaSep}), which are UV divergent and require a
regularization procedure. We will need to evaluate first the gap
equation for mass $M$, corresponding to the chirally broken phase. To
this end, we take the derivative of Eq.~(\ref{OmegaSep}) with respect
to $M$ to get
\begin{equation}
\frac{\partial\Omega}{\partial
  M}=\frac{\partial\Omega_{\text{finite}}}{\partial M}
+\frac{\partial\Omega_{T}}{\partial
  M}+\frac{\partial\Omega_{\text{Reg}}}{\partial M} . 
\end{equation}
In the derivatives of $\Omega_{\text{Reg}}$ we have divergent
integrands with the form 
\begin{eqnarray}
I_{M} & = & \int\frac{d^{3}p}{(2\pi)^{3}}\frac{1}{E}+\sum_{s=\pm1}
\int\frac{d^{3}p}{(2\pi)^{3}}\frac{1}{E}\frac{E+s\bar{\mu}}{E_{\Delta}^{s}}
\nonumber\\
& = & I_{M}^{a}+I_{M}^{b}.
 \label{Im}
\end{eqnarray}
To deal with the divergences of $I_{M}^{a}$, first of all we rewrite it
as 
\begin{eqnarray}
I_{M}^{a}&=&\int\frac{d^{3}p}{(2\pi)^{3}}\frac{1}{\sqrt{p^{2}+M^{2}}}
\nonumber \\ &=&
2\int_{-\infty}^{\infty}\frac{dp_{4}}{2\pi}\int\frac{d^{3}p}{(2\pi)^{3}}
\frac{1}{p_{4}^{2}+p^{2}+M^{2}}.
\label{Iqd4}
\end{eqnarray}
The identity equivalent to Eq.~(\ref{ident}) in the present case is 
\begin{eqnarray}
\lefteqn{ \frac{1}{p_{4}^{2}+p^{2}+M^{2}}  =
  \frac{1}{p_{4}^{2}+p^{2}+M_{0}^{2}} } \nonumber \\ && +
\frac{M_{0}^{2}-M^{2}}{(p_{4}^{2}+p^{2}+M_{0}^{2})(p_{4}^{2}+p^{2}+M^{2})},
\end{eqnarray}
which, iterated once, becomes
\begin{eqnarray}
\lefteqn{ \frac{1}{p_{4}^{2}+p^{2}+M^{2}}  =
  \frac{1}{p_{4}^{2}+p^{2}+M_{0}^{2}} } \nonumber \\ &&
+\frac{M_{0}^{2}-M^{2}}{(p_{4}^{2}+p^{2}+M_{0}^{2})^{2}}
\nonumber\\ & & +
\frac{(M_{0}^{2}-M^{2})^{2}}{(p_{4}^{2}+p^{2}+M_{0}^{2})^{2}(p_{4}^{2}+p^{2}+M^{2})}.
\end{eqnarray}
After performing the integrals indicated in~(\ref{Iqd4}), one gets 
\begin{eqnarray}
I_{M}^{a} & = &
I_{\text{quad}}+\frac{M_{0}^{2}-M^{2}}{2}I_{\text{log}}+I_{\text{fin,6}},
\label{intIma}
\end{eqnarray}
with 
\begin{equation}
I_{\text{fin,6}}=\frac{3}{4}\int\frac{d^{3}p}{(2\pi)^{3}}\int_{0}^{\infty}
\frac{t(M_{0}^{2}-M^{2})^{2}dt}{\sqrt{1+t}\left[(p^{2}+M_{0}^{2})t+p^{2}+M^{2}\right]^{\frac{5}{2}}},
\end{equation}
where we have used the same {}Feynman parametrization defined in
Eq.~(\ref{FeynParam}).

{}For $I_{M}^{b}$ we first write 
\begin{eqnarray}
I_{M}^{b} & = & \sum_{s=\pm1}\int\frac{d^{3}p}{(2\pi)^{3}}\frac{1}{E}
\frac{E+s\bar{\mu}}{E_{\Delta}^{s}} \nonumber \\ &=&
\sum_{s=\pm1}\int\frac{d^{3}p}{(2\pi)^{3}} \frac{1}{E_{\Delta}^{s}}
\nonumber \\ &+&
\sum_{s=\pm1}\int\frac{d^{3}p}{(2\pi)^{3}}\frac{s\bar{\mu}}{E}
\int_{-\infty}^{\infty}\frac{dp_{4}}{\pi}\frac{1}{p_{4}^{2}+(E_{\Delta}^{s})^{2}}.
\end{eqnarray}
Note that the first momentum integral in the right-hand side of the
above equation  is exactly $I_{\Delta}$ and determined in
Sec.~\ref{sec3}. To deal with the second momentum integral in the
above equation,  we can use the result Eq.~(\ref{r3it}) and write,
after performing the $p_{4}$ integration,
\begin{eqnarray}
\lefteqn{
  \frac{\bar{\mu}}{E}\int_{-\infty}^{\infty}\frac{dp_{4}}{\pi}\sum_{s=\pm1}
  \frac{s}{p_{4}^{2}+\left(E+s\bar{\mu}\right)^{2}+\Delta^{2}} }
\nonumber \\ && =
-\frac{4\bar{\mu}^{2}}{\left(p_{4}^{2}+p^{2}+M_{0}^{2}\right)^{2}}
-\frac{8A\bar{\mu}^{2}}{\left(p_{4}^{2}+p^{2}+M_{0}^{2}\right)^{3}}
\nonumber \\ & & +
\sum_{s=\pm1}\frac{s\bar{\mu}\left(A-2sE\bar{\mu}\right)^{3}}
    {E\left(p_{4}^{2}+p^{2}+M_{0}^{2}\right)^{3}\left[p_{4}^{2}
        +\left(E+s\bar{\mu}\right)^{2}+\Delta^{2}\right]}.  \nonumber
    \\
 \label{tempI7}
\end{eqnarray}
Then, we have 
\begin{eqnarray}
&&
  \int\frac{d^{3}p}{(2\pi)^{3}}\frac{\bar{\mu}}{E}\int_{-\infty}^{\infty}
  \frac{dp_{4}}{\pi}
  \sum_{s=\pm1}\frac{s}{p_{4}^{2}+\left(E+s\bar{\mu}\right)^{2}+\Delta^{2}}
  \nonumber \\ && =
  -2\bar{\mu}^{2}I_{\text{log}}-3A\bar{\mu}^{2}I_{\text{fin,1}}+I_{\text{fin,7}},
\label{intImb}
\end{eqnarray}
with, by using the same Feynman parametrization Eq.~(\ref{FeynParam})
again in the last line of Eq.~(\ref{tempI7}),
\begin{eqnarray}
\lefteqn{ I_{\text{fin,7}}
  =\frac{15}{16}\sum_{s=\pm1}\int\frac{d^{3}p}{(2\pi)^{3}}\int_{0}^{\infty}
  dt  \frac{t^2}{\sqrt{1+t}}} \nonumber \\ && \times
\frac{s\bar{\mu}\left(A-2sE\bar{\mu}\right)^{3}}{E
  \left[\left(p^{2}+M_{0}^{2}\right)t+\left(E+s\bar{\mu}\right)^{2}+
    \Delta^{2}\right]^{\frac{7}{2}}}.
\end{eqnarray}
In this way, collecting the results (\ref{intIma}) and (\ref{intImb})
in Eq.~(\ref{Im}), $I_{M}$ becomes 
\begin{eqnarray}
I_{M} & = & I_{M}^{a}+I_{M}^{b} \nonumber \\ & = &
3I_{\text{quad}}-\frac{\left(2\Delta^{2}-3M_{0}^{2}+3M^{2}\right)}{2}
I_{\text{log}}\nonumber
\\ & + &
\frac{3}{4}\left[A^{2}+\left(M^{2}-M_{0}^{2}\right)\bar{\mu}^{2}-
  4A\bar{\mu}^{2}\right]
I_{\text{fin,1}} \nonumber \\ &+ &
2I_{\text{fin,2}}+I_{\text{fin,6}}+I_{\text{fin,7}}.
 \label{ResIm}
\end{eqnarray}
Going back to the expression for $\Omega_{\text{Reg}}$ and recalling
all the definitions of Sec.\ref{sec2}, we rewrite it as
\begin{eqnarray}
\Omega_{\text{Reg}} & = & -4\int\frac{d^{3}p}{\left(2\pi\right)^{3}}
\left[\sqrt{\left(\sqrt{p^{2}+M^{2}}+\bar{\mu}\right)^{2}+\Delta^{2}}
  \right.  \nonumber \\ &+&
  \left. \sqrt{\left(\sqrt{p^{2}+M^{2}}-\bar{\mu}\right)^{2}+\Delta^{2}}\right.
  \nonumber \\ &  &
  +\left.\sqrt{p^{2}+M^{2}}-3\sqrt{p^{2}+M_{0}^{2}}\right].
\end{eqnarray}
Now, starting from the complete $M$ gap equation,  we perform an
integration in $M$ to obtain the thermodynamic potential, such that
\begin{equation}
\Omega=\int dM \frac{\partial\Omega}{\partial M} = \int d M
\left(\frac{\partial\Omega_{\text{finite}}}{\partial M}
+\frac{\partial\Omega_{T}}{\partial
  M}+\frac{\partial\Omega_{\text{Reg}}}{\partial M}\right).
\end{equation}
Note that once the integral in undefined, we obtain an integration
constant that has to be adjusted to obtain the same potential of the
TRS case, when the integrals in the MSS case are performed up to
$\Lambda$.  In this case, the numerical value of the expressions for
both schemes has to be exactly the same. To this end, we first
separate the normalized contribution of $r,g$, and $b$ quark colors,
namely,
\begin{eqnarray}
\Omega_{\text{Reg}} & = & \Omega_{r,g} + \Omega_{b}\nonumber\\ & = &
-4\int\frac{d^3k}{(2\pi)^3}\left(E_{\Delta}^+ + E_{\Delta}^-
-2E_0\right)   \nonumber \\ &-& 4\int\frac{d^3k}{(2\pi)^3}\left(E -
E_0\right),
\end{eqnarray}
where we have used the previous definitions of  $E_{\Delta}^{\pm} =
\sqrt{\left(E\pm\bar{\mu}\right)^{2}+\Delta^{2}}$  (remembering that
$E = \sqrt{p^2 + M^2}$)  and also $E_0 = \sqrt{p^2 + M_0^2}$. After
the $M$ integration and some  algebraic manipulations, we obtain
\begin{eqnarray}
\Omega_{r,g} & = & -4\int\frac{d^3k}{(2\pi)^3}\left(E_{\Delta}^+ +
E_{\Delta}^- -2E_0\right) \nonumber\\ & = & -4\bar{M}I_{\text{quad}}
-4\left(\Delta^2\bar{\mu}^2  - \frac{\bar{M}}{4}\right)I_{\text{log}}
\nonumber\\ & - &
4\int\frac{d^3k}{(2\pi)^3}\left[\left(\frac{\bar{M}^2}{4}  -
  \Delta^2\bar{\mu}^2\right)\frac{1}{E_0^3} \right.  \nonumber \\ &-&
  \left. \frac{\bar{M}}{E_0} - 2E_0 + E_{\Delta}^{+} +
  E_{\Delta}^{-}\right],
\end{eqnarray}
with the definition $\bar{M} = \Delta^2 + M^2 - M_0^2$ and, finally,
\begin{eqnarray}
\Omega_{b} & = & -4\int\frac{d^3k}{(2\pi)^3}\left(E - E_0\right)
\nonumber\\ & = & -2(M^2 - M_0^2)I_{\text{quad}} + \frac{(M^2 -
  M_0^2)^2}{2}I_{\text{log}}\nonumber\\ & - &
4\int\frac{d^3k}{(2\pi)^3}\left[E - E_0 - \frac{M^2 -M_0^2}{2E_0}  +
  \frac{(M^2 - M_0^2)^2}{8E_0^3}\right].  \nonumber \\
\end{eqnarray}
It is important to notice that the only divergent contributions,
$I_{\text{quad}}$ and $I_{\text{log}}$, already defined in
Sec.~\ref{sec3}, do not depend on medium contributions, only on the
vacuum mass $M_{0}$.

%%%%%%%%%%%%%%%%%%%%%%%%%%%%%%%%%%%%%%%%%%%%%%%%%%%%%%%%%%%%%%%%%%%
\section{The baryonic and individual densities}
\label{CalcDensities}

To obtain the baryon density $\rho_{B}$, we need to determine the total
density $\rho_{T}=\rho_{u}+\rho_{d}$ in the $SU(2)$ case. However, our
expression for the thermodynamic potential Eq.~(\ref{OmegaT0m0}) is
written in terms of $\bar{\mu}$ and $\delta\mu$. To rewrite it in
terms of the $u$ and $d$ quark chemical potentials, we first write
\begin{eqnarray}
\Omega_{T=0} & = &
\Omega_{0}-\frac{\mu_{e}^{4}}{12\pi^{2}}+\frac{\Delta^{2}}{4G_{d}}
-4\int_0^{\Lambda}\frac{dp}{2\pi^2}p^3-\frac{\mu_{ub}^{4}}{12\pi^{2}}
\nonumber \\ &-&\frac{\mu_{db}^{4}}{12\pi^{2}}+\Omega_{\mu},
\end{eqnarray}
where we identify 
\begin{eqnarray}
\Omega_{\mu}&=&-4\int\frac{d^{3}p}{\left(2\pi\right)^{3}}\left(
E_{\Delta}^{+} +E_{\Delta}^{-}\right) \nonumber \\ &-&
2\theta\left(\delta\mu-\Delta\right)\int_{\mu^{-}}^{\mu^{+}}
\frac{dp}{\pi^{2}}p^{2}\left(\delta\mu-E_{\Delta}^{-}\right).
\label{Omegamu}
\end{eqnarray}
Note that $\Omega_{\mu}$ refers to the double degenerate modes {[}see
  Eqs.~(\ref{Eub})-(\ref{Erg}){]},  so one can write 
\begin{eqnarray}
\Omega_{\mu}\left(\Delta,\bar{\mu},\delta\mu\right)  & = &
\frac{1}{2}\Omega_{\mu}
\left(\Delta,\frac{\mu_{dg}+\mu_{ur}}{2},\frac{\mu_{dg}-\mu_{ur}}{2}\right)
\nonumber \\ &+&
\frac{1}{2}\Omega_{\mu}
\left(\Delta,\frac{\mu_{dr}+\mu_{ug}}{2},\frac{\mu_{dr}-\mu_{ug}}{2}\right).
\nonumber \\
\label{Wdeg}
\end{eqnarray}
In this way, we can write $\rho_{u}=\rho_{ur}+\rho_{ug}+\rho_{ub}$.
Due to the gauge choice, we have $\rho_{ur}=\rho_{ug}$. {}For the
quark $u$ we have 
\begin{eqnarray}
\rho_{ur}&=&\rho_{ug}  =
-\frac{\partial\Omega_{\mu}\left(\Delta,\bar{\mu},\delta\mu\right)}
{\partial\mu_{ur}}
\nonumber \\ & = &
-\frac{1}{4}\left[\frac{\partial\Omega_{\mu}
    \left(\Delta,\bar{\mu},\delta\mu\right)}
  {\partial\bar{\mu}}-\frac{\partial\Omega_{\mu}
    \left(\Delta,\bar{\mu},\delta\mu\right)}
  {\partial\left(\delta\mu\right)}\right],
\label{rhoUrg}
\end{eqnarray}
and in the blue direction,
\begin{eqnarray}
\rho_{ub} & = &
-\frac{\partial\Omega_{T=0}}{\partial\mu_{ub}}=\frac{\mu_{ub}^{3}}{3\pi^{2}}.
\label{rhoUb}
\end{eqnarray}
{}For the quark $d$, we have $\rho_{dr}=\rho_{dg}$, such that
\begin{eqnarray}
\rho_{dr}=\rho_{dg} & = &
-\frac{\partial\Omega_{\mu}\left(\Delta,\bar{\mu},\delta\mu\right)}
{\partial\mu_{dr}}\nonumber \\ & = &
-\frac{1}{4}\left[\frac{\partial\Omega_{\mu}
    \left(\Delta,\bar{\mu},\delta\mu\right)}
  {\partial\bar{\mu}}+\frac{\partial\Omega_{\mu}
    \left(\Delta,\bar{\mu},\delta\mu\right)}
  {\partial\left(\delta\mu\right)}\right], \nonumber \\
 \label{rhoDrg}
\end{eqnarray}
while in the blue direction,
\begin{equation}
  \rho_{db}=-\frac{\partial\Omega_{T=0}}{\partial\mu_{db}}=
  \frac{\mu_{db}^{3}}{3\pi^{2}}.
\label{rhoDb}
\end{equation}

Evaluating the derivatives in Eqs.~(\ref{rhoUrg}) and (\ref{rhoDrg}),
we obtain
\begin{eqnarray}
\frac{\partial\Omega\left(\Delta,\bar{\mu},\delta\mu\right)}{\partial\bar{\mu}}
& = & -4I_{8}^{i}-4\bar{\mu}I_{\Delta}^{i} \nonumber \\ &-&
2\theta\left(\delta\mu-\Delta\right)
\int_{\mu^{-}}^{\mu^{+}}\frac{dp\,p^{2}}{\pi^{2}}
\left(\frac{p-\bar{\mu}}{E_{\Delta}^{-}}\right), \nonumber \\
\label{pWpmubar}
\end{eqnarray}
\begin{eqnarray}
\lefteqn{
  \frac{\partial\Omega\left(\Delta,\bar{\mu},\delta\mu\right)}
       {\partial\left(\delta\mu\right)}
} \nonumber \\ & & =
-\frac{4\sqrt{\delta\mu^{2}-\Delta^{2}}}{3\pi^{2}}\left(\delta\mu^{2}
-\Delta^{2}+3\bar{\mu}^{2}\right)\theta\left(\delta\mu-\Delta\right).
\nonumber \\
\label{pWpdmu}
\end{eqnarray}
{}Finally, the expressions for the $\rho_{u}$ and $\rho_{d}$ densities
become
\begin{eqnarray}
\rho_{u} & = &
2I_{8}^{i}+2\bar{\mu}I_{\Delta}^{i}+\theta\left(\delta\mu-\Delta\right)
\int_{\mu^{-}}^{\mu^{+}}\frac{dp}{\pi^{2}}p^{2}\left(\frac{p-\bar{\mu}}
    {E_{\Delta}^{-}}\right)
\nonumber \\ & - & \frac{2\sqrt{\delta\mu^{2}-\Delta^{2}}}{3\pi^{2}}
\left(\delta\mu^{2}-\Delta^{2}+3\bar{\mu}^{2}\right)
\theta\left(\delta\mu-\Delta\right) \nonumber \\ &+&
\frac{\mu_{ub}^{3}}{3\pi^{2}},
 \label{rhoufin}
\end{eqnarray}
and
\begin{eqnarray}
\rho_{d} & = &
2I_{8}^{i}+2\bar{\mu}I_{\Delta}^{i}+\theta\left(\delta\mu-\Delta\right)
\int_{\mu^{-}}^{\mu^{+}}\frac{dp}{\pi^{2}}p^{2}\left(\frac{p-\bar{\mu}}
    {E_{\Delta}^{-}}\right)
\nonumber \\ & + & \frac{2\sqrt{\delta\mu^{2}-\Delta^{2}}}{3\pi^{2}}
\left(\delta\mu^{2}-\Delta^{2}+3\bar{\mu}^{2}\right)
\theta\left(\delta\mu-\Delta\right) \nonumber \\ &+&
\frac{\mu_{db}^{3}}{3\pi^{2}},
 \label{rhodfin}
\end{eqnarray}
where $I_{8}^{i}$ and $I_{\Delta}^{i}$ were defined previously for
each scheme and given by Eqs.~(\ref{IdTRS}),~(\ref{IdMSS}),
~(\ref{I8TRS}), and~(\ref{I8MSS}).

%%%%%%%%%%%%%%%%%%%%%%%%%%%%%%%%%%%%%%%%%%%%%%%%%%%%%%%%%%%%%%%%%%%%%%%%%%%%%

\end{document}